\input harvmac
\input psfig
\newcount\figno
\figno=0
\def\fig#1#2#3{
\par\begingroup\parindent=0pt\leftskip=1cm\rightskip=1cm\parindent=0pt
\global\advance\figno by 1
\midinsert
\epsfxsize=#3
\centerline{\epsfbox{#2}}
\vskip 12pt
{\bf Fig. \the\figno:} #1\par
\endinsert\endgroup\par
}
\def\figlabel#1{\xdef#1{\the\figno}}
\def\encadremath#1{\vbox{\hrule\hbox{\vrule\kern8pt\vbox{\kern8pt
\hbox{$\displaystyle #1$}\kern8pt}
\kern8pt\vrule}\hrule}}

\overfullrule=0pt
\def\RP{{\bf RP}}
%
\def\underarrow#1{\vbox{\ialign{##\crcr$\hfil\displaystyle
 {#1}\hfil$\crcr\noalign{\kern1pt\nointerlineskip}$\longrightarrow$\crcr}}}
%
\def\tilde{\widetilde}

\def\Z{{\bf Z}}

\def\S{{\bf S}}
\def\R{{\bf R}}

\font\zfont = cmss10 
\font\litfont = cmr6

\def\bigone{\hbox{1\kern -.23em {\rm l}}}
\def\ZZ{\hbox{\zfont Z\kern-.4emZ}}
\def\half{{\litfont {1 \over 2}}}

\Title{hep-th/9805112, IASSNS-HEP-98-42}
{\vbox{\centerline{Baryons And Branes In Anti de Sitter Space}}}
\bigskip
\centerline{Edward Witten}
\smallskip
\centerline{\it School of Natural Sciences, Institute for Advanced Study}
\centerline{\it Olden Lane, Princeton, NJ 08540, USA}\bigskip

\medskip
\def\N{{\cal N}}
\noindent
In the mapping from four-dimensional gauge theories to string
theory  in $AdS$ space, many features of gauge
theory can be described by branes wrapped in different ways on
$\S^5$, $\RP^5$, or subspaces therefore.  These include a baryon
vertex coupling $N$ external charges in the fundamental representation
of $SU(N)$, a bound state of $k$ gluons in $SO(2k)$ gauge theory,
strings coupled to external charges in the spinor representation
of the gauge group, and domain walls across which the low energy gauge group
changes.
\Date{May, 1998}
\newsec{Introduction}

\def\RP{{\bf RP}}
It has been argued \ref\malda{J. Maldacena, ``The Large $N$ Limit
Of Superconformal Field Theories And Supergravity,'' hep-th/9711200.} 
that four-dimensional
$\N=4$ super Yang-Mills theory, with gauge group $SU(N)$, is equivalent
to Type IIB superstring theory
 on $AdS_5\times \S^5$ (where $AdS_5$ is
five-dimensional Anti-de Sitter space, and there are $N$ units of five-form
flux on the five-sphere
$\S^5$). This fascinating subject has been developed in many directions.
Precise recipes for computing correlation functions
of local operators in this framework have been presented in 
\nref\kleb{S. S. Gubser, I. R. Klebanov, and A. M. Polyakov,
``Gauge Theory Correlators From Non-Critical String Theory,''
hep-th/9802109.}\nref\witten{E. Witten, ``Anti de Sitter Space And Holography,''
hep-th/9802150.}\refs{\kleb,\witten}; there likewise
are precise recipes for computing correlation functions involving Wilson
loop operators 
\nref\maldacena{J. Maldacena, ``Wilson Loops In Large $N$ Field Theories,''
hep-th/9803002.}
\nref\reyyee{S.-J. Rey and J. Yee, ``Macroscopic Strings As Heavy
Quarks Of Large $N$ Gauge Theory And Anti-de SItter
Supergravity,'' hep-th/9803001.}\refs{\maldacena,\reyyee}.
A similar treatment can be given for models with reduced
supersymmetry that are obtained by an orbifolding operation in which
$\S^5$ is replaced by $\S^5/\Gamma$, with $\Gamma$ a finite group
\ref\silverstein{S. Kachru and E. Silverstein, ``4d Conformal Field Theories
And Strings On Orbifolds,'' hep-th/9802183.}.  
\nref\oz{O. Aharony, Y. Oz, and Z. Yin, ``$M$ Theory On $AdS_p\times \S^{11-p}$
and Superconformal Field Theories,'' hep-th/9803051.}
Of particular importance
in the present paper, the gauge group $SU(N)$ can be replaced by
$SO(N)$ or $Sp(N/2)$ by an orientifolding operation 
in which $\S^5$ is replaced by $\RP^5=\S^5/\Z_2$
\nref\kaku{Z. Kakushadze, ``Gauge Theories From Orientifolds And
Large $N$ Limit,'' hep-th/9803214, ``On Large $N$ Gauge Theories
From Orientifolds,'' hep-th/9804184.}
\nref\fayy{A. Fayyazuddin and M. Spali\'nski, ``Large $N$ Superconformal
Gauge Theories And Supergravity Orientifolds,'' hep-th/9805096.}
 (analogous orientifolds
 in eleven dimensions
were discussed in \oz; such orbifolds in ten dimensions were discussed
in \kaku\ and related explicitly to supergravity in \fayy).

\def\Pf{{\rm Pf}}
The present paper began with the following question.  Since the
$AdS_5\times \S^5$ theory encodes an $SU(N)$ gauge theory, rather than $U(N)$,
should there not be a baryon vertex?  In other words, should there not
be finite energy configurations with $N$ external quarks, roughly
in parallel with the external quark-antiquark configurations studied in
\refs{\maldacena,\reyyee}?  In section 2, we will construct such
a baryon vertex.  It has a simple interpretation; it is obtained
by wrapping a fivebrane over $\S^5$!  

Finding a baryonic vertex in the $\N=4$ theory does not mean that
that theory has baryonic particles, or operators.  Baryonic particles
would appear in a theory that has
dynamical quark fields
(that is, fields transforming in the fundamental representation of $SU(N)$);
in their absence, we get only a baryonic vertex, a gauge-invariant
coupling of $N$ external charges.
Introducing dynamical quark fields would require breaking some
supersymmetry.  As an alternative
route to studying an object somewhat like a baryonic particle, we can
replace the gauge group $SU(N)$ with $SO(N)$, without breaking any 
supersymmetry.  Take $N$ to be an even number, $N=2k$.
 $SO(2k)$ gauge theory admits a gauge-invariant configuration
of $k=N/2$ gauge bosons.  In fact, if $\Phi_{ab}, \,a,b=1,\dots,2k$ is an 
antisymmetric second rank tensor, transforming in the adjoint representation
of $SO(2k)$, then the ``Pfaffian'' $\Pf(\Phi)=(1/k!)\epsilon^{a_1a_2\dots
a_{2k}}\Phi_{a_1a_2}\dots \Phi_{a_{2k-1}a_{2k}}$ is an irreducible
gauge-invariant polynomial of order $N/2$.  We will call such operators
Pfaffian operators, and the particles they create Pfaffian particles.

The Pfaffian  particle has long presented a puzzle for the general
understanding of the $1/N$ expansion of gauge theories.\foot{Roughly
such questions about $N$ dependence motivated early pre-$D$-brane
work (see for example \ref\gre{M. B. Green, ``Space-Time Duality
And Dirichlet String Theory,'' Phys. Lett. {\bf B266} (1991) 325,
``Point-Like States For Type II Superstrings,'' Phys. Lett. {\bf B329} 
(1994) 434.}) and early
conjectures \ref\shenker{S. H. Shenker, ``The Strength Of Non-perturbative
Effects In String Theory,'' in Carg\'ese 1990, {\it Random Surfaces
And Quantum Gravity}.} 
about the structure of nonperturbative
corrections in string theory.}  It is suspected 
\ref\thooft{G. 't Hooft, ``A Planar Diagram Theory For Strong
Interactions,'' Nucl. Phys. {\bf B72} (1974) 461.} 
that Yang-Mills theory with $SU(N)$, $SO(N)$,
or $Sp(N)$ gauge group has a large $N$ limit as a closed string theory,
with an effective string coupling
constant $\lambda\sim 1/N$.   If elementary quarks are added, 
it is believed that there are also open strings (describing mesons),
with an open string coupling constant
$\lambda'\sim 1/\sqrt N$.
Baryons of an $SU(N)$ theory with dynamical quarks are $N$-quark bound
states, which one would expect to have masses of order $N$.  As $N\sim
1/(\lambda')^2$, such states can be interpreted as solitons in the open
string sector \ref\oldwitten{E. Witten, ``Baryons In The $1/N$ Expansion,''
Nucl. Phys. {\bf B160} (1979) 57. }.

The Pfaffian particle of an $SO(N)$ gauge theory without quarks, 
being a bound state
of $N/2$ gluons, is intuitively expected to have a mass of order $N$.
This particle cannot be interpreted as a soliton because, in terms of
the closed string coupling $\lambda$, its mass is of order $1/\lambda$,
not $1/\lambda^2$.  Given the modern understanding of $D$-branes \ref\polch{
J. Polchinski, ``Dirichlet Branes And Ramond-Ramond Charges,'' Phys. Rev.
Lett. {\bf 75} (1995) 4724, hep-th/9510017.}, one might wonder if the Pfaffian
particle of $SO(N)$ is a $D$-brane.
\foot{Likewise, might the baryon of an $SU(N)$ theory with dynamical
quarks be
a $D$-brane rather than an open string soliton?  This question is
not quite well-defined because the $SU(N)$ theory with quarks has
open strings.  In theories with open strings,
$D$-branes and open string solitons can describe the same objects,
as in the case of Type I fivebranes/Yang-Mills instantons
\ref\ewitten{E. Witten, ``Small Instantons In String Theory,'' Nucl. Phys.
{\bf B460} (1996) 541.}.
Which description is more useful can depend on the circumstances.}

This  question can  be addressed by comparing
$SO(N)$ gauge theory to string theory on $AdS_5\times \RP^5$.  
With this aim,
 we describe  in section 3 the basic rules for wrapping branes on
$\RP^5$.  We explain the ``discrete torsion'' that distinguishes
the $SO(2k)$, $SO(2k+1)$, and $Sp(k)$ theories, and its relation to
Montonen-Olive duality.  We also describe some important topological
restrictions on brane wrapping.

\nref\otherwitten{E. Witten, ``Anti de Sitter Space, Thermal Phase Transition,
And Confinement In Gauge Theories,'' hep-th/9803131.}
We then go on  in section 4 to discuss the gauge theory interpretation
of various types of wrapped branes. In section 4.1, we
show that in $SO(2k)$ gauge theory, interpreted in terms of $AdS_5\times 
\RP^5$,  there  is a natural
candidate for the Pfaffian  -- a threebrane wrapped on
an $\RP^3$ subspace of $\RP^5$.\foot{As explained at the end of
 section 4.1, the wrapped threebrane is really related to a Pfaffian
operator of the boundary theory rather than a  Pfaffian particle; it would
be related to a Pfaffian particle after a suitable perturbation that
breaks conformal invariance.}
We show that a threebrane cannot be so wrapped in $SO(2k+1)$
or $Sp(k)$ -- as expected, since these groups do not have gauge-invariant
Pfaffians.  For $SO(2k+1)$, we show that one can have a  wrapped
threebrane with an attached string; we explain the meaning of this
object in gauge theory.

The rest of section 4 is devoted to other types of brane wrapping.
 In section 4.2, we consider strings obtained by wrapping
a fivebrane on an $\RP^4$ subspace of $\RP^5$. We argue that such
strings can be used to compute, in $SO(2k)$ or $SO(2k+1)$
gauge theory, the expectation values  
of Wilson lines in the spinor representation of the gauge
group.  (We also show that
such wrapped branes are not possible in $Sp(k)$ gauge theory -- as expected
since $Sp(k)$ has no spinor representation.)
In section 4.3, we consider threebranes on $AdS_5$ -- both the unwrapped
Type IIB threebrane and additional threebranes made by wrapping a fivebrane
on $\RP^2\subset \RP^5$.  We argue that such threebranes behave as domain
walls, with the property that the gauge group jumps when one crosses one.
In section 4.4, we consider the baryon vertex -- the antisymmetric
coupling of $N$ external quarks -- in the orthogonal and symplectic gauge
theories.
Finally, in section 4.5, we reexamine the $-1$-branes of $AdS$ space
-- which were recently discussed as Yang-Mills instantons \ref\banks{T.
Banks and M. B. Green, ``Nonperturbative Effects in $AdS_5\times\S^5$
String Theory and $D=4$ SUSY Yang-Mills,'' hep-th/9804170.}.

\newsec{The Baryon Vertex In $SU(N)$ Gauge Theory}

First we recall how external quarks in $\N=4$ super Yang-Mills
are described in terms of Type IIB on $AdS_5\times \S^5$ 
\refs{\maldacena,\reyyee}.
With Lorentz signature, the boundary of $AdS_5$ is $\S^3\times \R$,
where $\R$ is the ``time'' direction; $\S^3\times \R$ is a universal
cover of the conformal compactification of Minkowski space.
External quarks are regarded as endpoints of  strings
in $AdS$ space.  Thus, to compute the energy for a
time-independent configuration with an external quark at a point
$x\in \S^3$ and an external antiquark at $y\in \S^3$, one considers
configurations in which a string inside
the $AdS$ space  connects the boundary points $x$ and $y$.

The strings in question are elementary Type IIB superstrings
if the external charges are electric charges, that is particles in the
fundamental representation of the gauge group $SU(N)$.  External
monopoles would be boundaries of $D$-strings, and external charges
of type $(p,q)$ are boundaries of $(p,q)$ strings. 

\bigskip
\centerline{\vbox{\hsize=4in\tenpoint
\centerline{\psfig{figure=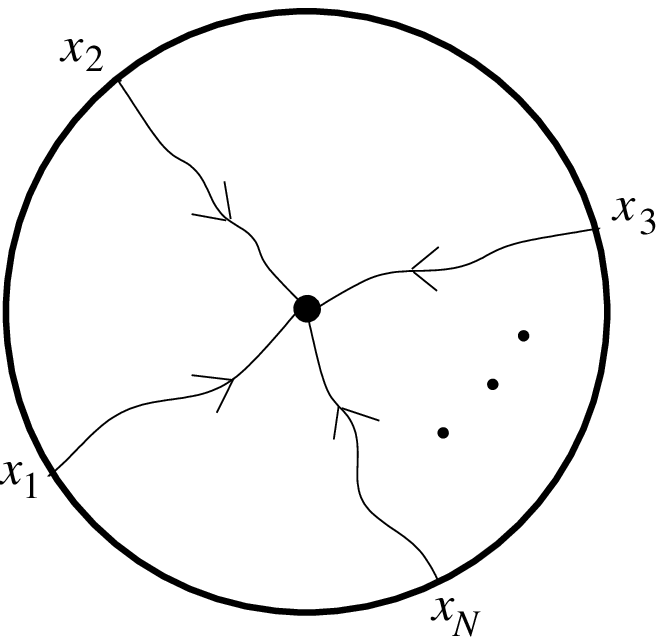}}
\vglue.4in
Fig. 1.  $N$ elementary strings attached to points $x_1,x_2,\dots ,x_N$
on the boundary of $AdS$ space and joining at a baryon vertex in the interior.}}

As promised in the introduction, we now want to find a ``baryon''
vertex connecting $N$ external quarks, with their color wave functions
combined together
by an $N^{th}$ order antisymmetric tensor of $SU(N)$.
For this, we place external
quarks at boundary points $x_1,x_2,\dots,x_N$.  We consider a configuration
in which each of the boundary points is the endpoint of an elementary
superstring in $AdS_5\times \S^5$, with all $N$ strings oriented in the same
way.  We want, as in figure 1, to find a ``baryon vertex,''
where these $N$ strings can somehow terminate in the interior of
$AdS_5\times \S^5$.

We claim that the baryon vertex is simply a wrapped fivebrane -- 
i.e., a fivebrane whose world-volume is   $ \S^5\times \R$, with $\R$ 
a timelike curve in $AdS_5$.  With a suitable choice of ``time'' coordinate
in $AdS_5$, this is the worldvolume of a static fivebrane wrapped
on $\S^5\times Q$, where $Q$ is a point in a time zero slice of 	
$AdS_5$.  Assuming that the strings that are to be joined
at the baryon vertex are elementary strings, we will build the baryon
vertex from a $D$-fivebrane; the baryon vertex for $N$ external
charges of the same type $(p,q)$ would be similarly made by 
wrapping a $(p,q)$ fivebrane on $\S^5$.  For definiteness, in what
follows, we consider the case of elementary strings and $D$-fivebranes.

The reason that the wrapped fivebrane is a baryon vertex is the following.
In Type IIB superstring theory, there is a self-dual five-form field $G_5$.
The $AdS_5\times \S^5$ compactification which is related to $SU(N)$ gauge
theory has $N$ units of five-form flux on $\S^5$:
\eqn\nobo{\int_{\S^5}{G_5\over 2\pi}=N.}
On the $D$-fivebrane worldvolume, there is a $U(1)$ gauge field $a$.
It couples to $G_5$ by a coupling
\eqn\loobo{\int_{\S^5\times \R}a\wedge {G_5\over 2\pi}.}
Because of this coupling and \nobo, the $G_5$ field contributes
$N$ units of $a$-charge.  Since the total charge of a $U(1)$ gauge
field must vanish in a closed universe, there must be $-N$ units
of charge from some other source.

Such a source is an elementary string ending on the fivebrane.
As in \ref\stromfun{A. Strominger, ``Open $p$-Branes,'' Phys. Lett.
{\bf B383} (1996) 44.},
the endpoint of an elementary string that ends on the fivebrane
is electrically charged with respect to the $a$-field, with a charge
that is $+1$ or $-1$ depending on the orientations of the string
and fivebrane.  To cancel the $G_5$ contribution to the $a$-charge,
we need $N$ strings, all oriented in the same way, ending on the
fivebrane.  The fivebrane is thus a baryon or antibaryon vertex, depending
on its orientation.

We have thus  provided evidence that the gauge theory on $\S^3\times \R$
that is dual to Type IIB on $AdS_5\times \S^5$ has the property
that  it is possible to form a gauge-invariant combination of $N$ quarks,
that is, of $N$ particles in the fundamental representation of the gauge
group.  This is in agreement with the fact that the gauge group is
believed to be $SU(N)$ (and not $U(N)$, for example).  But group theory
predicts more.  In $SU(N)$
gauge theory, the gauge-invariant combination of $N$ elementary quarks
is completely antisymmetric.  How do we see the antisymmetry
in $AdS_5\times \S^5$ string theory?
 
The antisymmetry means that elementary strings that connect 
a $D$-fivebrane to the boundary of $AdS$ space behave as fermions.
Since the boundary of $AdS$ is only at a finite distance from a conformal
point of view, to describe a string that stretches to the boundary of
$AdS$ space, we need a boundary condition at spatial infinity.
We will now show that with certain natural boundary conditions,
the strings stretching to the boundary in fact behave as fermions,
giving the antisymmetry of the baryon vertex.

\bigskip
\centerline{\vbox{\hsize=4in\tenpoint
\centerline{\psfig{figure=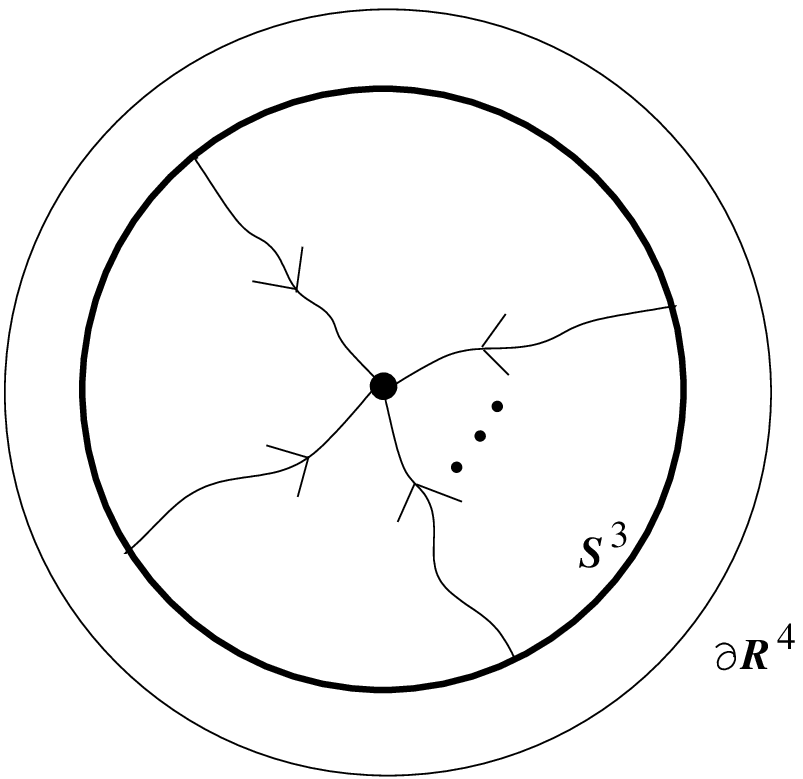}}
\vglue.4in
Fig. 2. This differs from figure 1 in that now a threebrane wraps
over a large three-sphere in $AdS$ space, and the $N$ elementary strings
terminate on the threebrane rather than the boundary.  In the figure,
the boundary of $AdS$ space (a spatial section of which is topologically
${\bf R}^4$) is denoted $\partial\R^4$, and the spatial section
of the threebrane is denoted $\S^3$.}}

Natural boundary conditions are suggested by the original argument
\refs{\maldacena,\reyyee} for regarding
elementary quarks on $\S^3\times \R$ as boundaries of strings
in $AdS$ space.  In this argument, strings with an endpoint 
 at infinity are regarded as limiting cases of strings with an
 endpoint on a three-brane that is, in a suitable sense, near to infinity.
Thus, a time zero section of $AdS_5$ is a copy of $\R^4$, so a time zero
section of $AdS_5\times \S^5$ is a copy of $\R^4\times \S^5$.  
We consider (figure 2)
a static threebrane whose world-volume, at time zero, is a subspace of
$\R^4\times \S^5$ of the
form $\S^3\times R$, with $\S^3$ a sufficiently large three-sphere near
infinity in $\R^4$ and  $R$ a point in $\S^5$.
Our $N$ strings thus connect the threebrane on $\S^3\times R$ with
the fivebrane on $Q\times \S^5$.

Now, the three-manifolds $\S^3\times R$ and $Q\times \S^5$ are ``linked''
in the nine-manifold $\R^4\times \S^5$ -- they have linking number
$\pm 1$, depending on orientation.  It has been seen in \ref\hanany{A. 
Hanany and E. Witten, ``Type IIB Superstrings, BPS Monopoles,
And Three-Dimensional Gauge Dynamics,'' Nucl. Phys. {\bf B492} (1997) 152.}
that such linked branes, under certain conditions, are connected
by elementary strings; the argument is a close cousin of the argument
by which we deduced above that the wrapped fivebrane in $AdS_5\times \S^5$
is a baryon vertex.  As part of one explanation of the linking phenomenon,
it has been observed \ref\green{C. I. Bachas, M. R. Douglas, and
M. B. Green, ``Anomalous Creation Of Branes,'' hep-th/9705074,
``$(8,0)$ Quantum Mechanics And Symmetry Enhancement
In Type $I'$ Superstrings,'' hep-th/9712086.} that the ground
state of a  string stretching between linked $D$-branes
is fermionic and nondegenerate.\foot{Instead of linked $D$-branes,
one can take any transverse $D$-branes whose total dimensions add up to
eight; the picture looks the same locally.
If space is taken to be the nine-manifold
$\R^9$ with coordinates $x^1,\dots,x^9$, the  $D$-branes can be
 a threebrane located at
$x^4=x^5=\dots =x^9=0$ and a fivebrane located at $x^1=x^2=x^3=0$,
$x^4=b$, where our present problem corresponds to $b\not= 0$, 
but in section 4.2 we will consider a case with $b=0$.  
The ground state of a string connecting these branes is fermionic because
(given the boundary conditions at the two ends) the ground state energy
in the Neveu-Schwarz sector is positive while in the Ramond sector it
is zero.  The ground state is nondegenerate because there are no fermion
zero modes in the Ramond sector.}
This is just right to make the baryon vertex antisymmetric, so
if the boundary conditions are the ones implied in \refs{\maldacena,\reyyee},
the baryon vertex is completely antisymmetric under permutation of the $N$
strings.  

Let us estimate the energy of the baryon vertex 
in the 't Hooft limit (the string coupling $\lambda$ going
to zero, and $N$ to infinity, with $\lambda N$ fixed).
Since the $D$-brane tension is of order $1/\lambda$, that is of order $N$,
and the volume of $\S^5$ remains finite for $N\to\infty$,
the baryon vertex for $N$ external electric charges has an energy of
order $N$, as expected for baryons in the large $N$ limit of QCD \oldwitten.  
For $N$ external magnetic
charges, the energy of the baryon vertex is controlled by the tension
of an NS fivebrane, and has an extra factor of $1/\lambda$ or $N$; this
factor seems natural, as  magnetic charges
are boundaries of $D$-strings, whose tension is $1/\lambda$ or $N$ times
the tension of the elementary strings.  In each case, the energy of
the baryon vertex is comparable to the energy of the $N$ strings that
are attached to it, and hence a dynamical study of baryonic states
would involve balancing these two energies.

\newsec{Orthogonal and Symplectic Gauge Groups}

\subsec{The $\RP^5$ Orientifold}

By considering $N$ parallel threebranes in $\R^{10}$, one gets
a $U(N)$ gauge theory.  Upon consideration of the associated supergravity
solution and its near-horizon geometry, one gets the $AdS_5\times \S^5$
description of $SU(N)$ gauge theory.

To obtain $SO(N)$ or (for even $N$) $Sp(N/2)$ gauge symmetry, one can
instead consider $N$ parallel threebranes at an orientifold threeplane.
In other words, one replaces $\R^{10}$ by $\R^4\times (\R^6/\Z_2)$ 
(with the $\Z_2$ acting by sign change on all six coordinates of $\R^6$)
and places the threebranes at the singularity in $\R^6/\Z_2$.
The orientifolding operation replaces
a small      sphere around the origin in $\R^6$
by a copy of $\RP^5=\S^5/\Z_2$.\foot{We recall that $\RP^n$ -- real projective
$n$-space --  is the quotient of the sphere $x_1^2+x_2^2+\dots +x_{n+1}^2=1$
by the $\Z_2$ symmetry $x_i\to -x_i$.} So it
replaces the $\S^5$ factor in the
near-horizon geometry by $\RP^5$.  So one is led 
(see \refs{\oz,\kaku,\fayy}) to suspect that $\N=4$ super Yang-Mills theory
with orthogonal or symplectic gauge group can be described in terms of
Type IIB superstring theory on an $AdS_5\times \RP^5$ orientifold.

The statement that the $AdS_5\times \RP^5$ model is
an orientifold means that in going around a noncontractible loop
in the target space, the orientation of the string worldsheet is
reversed.  A formal way to say this is the following.  Let $x$
be the generator of $H^1(\RP^5,\Z_2)$, which is isomorphic to $\Z_2$. 
Let $\Sigma$ be a string
world-sheet (a closed and not necessarily orientable two-dimensional
surface), and let $w_1(\Sigma)\in H^1(\Sigma,\Z_2)$ be the obstruction
to its orientability.
Then we consider only maps $\Phi:\Sigma\to AdS_5\times \RP^5$
such that $\Phi^*(x)=w_1(\Sigma)$.

Since the $\S_2$ action on $\S^5$ is free, the $AdS_5\times \S^5/\Z_2$ 
orientifold
has no orientifold fixed points and thus no
open string sector.  There also is no ``winding sector'' 
consisting of closed strings wrapped around a non-contractible loop in
$\RP^5$.  (The latter statement is a general property of orientifolds,
whether the $\Z_2$ action is free or not; we recall the reason at the beginning
of section 3.3.)
The spectrum of the model, for weak coupling, is just
the $\Z_2$-invariant part of the spectrum of the
 $AdS_5\times \S^5$ compactification.

The interactions are, however, different, because in the orientifold
the string worldsheet $\Sigma$ need not be orientable.  A
basic case of an unorientable closed world-sheet is $\Sigma=\RP^2$,
which we can identify as the quotient of the two-sphere $x_1^2+x_2^2+x_3^2=1$
by the overall sign change $x_i\to -x_i$.  A typical map $\Phi:\RP^2\to \RP^5$
that obeys the orientifold constraint $\Phi^*(x)=w_1(\RP^2)$ is the embedding
\eqn\hunflo{(x_1,x_2,x_3)\to (x_1,x_2,x_3,0,0,0).}
For large $N$, the leading deviation of the $AdS_5\times \RP^5$ theory
from the $\Z_2$-invariant part of the $AdS_5\times \S^5$ theory
comes from such $\RP^2$ worldsheets.\foot{The $\RP^2$
contribution dominates
for large $N$ over the one-loop  contribution which obviously -- since
the $AdS_5\times \S^5$ theory has additional intermediate states
relative to $AdS_5\times \RP^5$ -- distinguishes the two theories.  The
$\RP^2$ amplitude is of order $\lambda$ relative to the $\S^2$ contribution
($\lambda$ being the string coupling constant), while the one-loop
contribution is of relative order $\lambda^2$.}

The structure just described agrees beautifully with expectations from
gauge theories.  In the framework of 't Hooft \thooft,  
$SO(N)$ and $Sp(N/2)$ gauge theories should be described for large
$N$ by the ``same'' string theory that governs the $SU(N)$ theory,
except that the strings, while still closed, are unoriented.
This conclusion  follows by analyzing, as in 
\ref\cicuta{G. M.  Cicuta, ``Topological Expansion For $SO(N)$
and $Sp(2N)$ Gauge Theories,'' Lett. Nuovo Cim. {\bf 35} (1982) 87.},
the Riemann surfaces that are built from Feynman diagrams
in the $SO(N)$ and $Sp(N/2)$ cases.
Hence the spectra of the $SO(N)$ and $Sp(N/2)$ theories should be obtained
from the spectrum of the $SU(N)$ theory by extracting the part invariant
under an orientifold projection.  The interactions of the $SO(N)$ and $Sp(N/2)$
gauge theories differ in perturbation theory 
from those of $SU(N)$, with the leading difference
for large $N$
coming from worldsheets (that is, Feynman diagrams)
with $\RP^2$ topology.  (See \cicuta\ for a detailed description of
Feynman diagrams with $\RP^2$ topology.)  Moreover, by analyzing
the Feynman diagrams, one can show that  the
$SO(N)$ and $Sp(N/2)$ gauge theories differ from each other essentially just
by the sign
of the $\RP^2$ contribution.\foot{In \nref\mkrt{R. L. Mkrtchyan,
``The Equivalence Of $Sp(2N)$ and $SO(2N)$ Gauge Theory,'' Phys. Lett.
{\bf 105B} (1981) 174.}
\nref\cvit{P. Cvitanovic and A. D. Kennedy, ``Spinors In Negative Dimensions,''
Physica Scripta {\bf 26} (1982) 5.}
\refs{\mkrt,\cvit}, it is shown that Feynman diagrams of $Sp(N/2)$
are obtained from those of $SO(N)$ via the simple transformation $N\to -N$.  
Since a diagram describing a Riemann surface of genus $g$ glued to
$s $ copies of $\RP^2$ is of order $N^{2-2g-s}$, the effect of $N\to -N$
is precisely to include a factor of $-1$ for each $\RP^2$.}
This last statement is in accord with
a feature of string theory on $AdS_5\times \RP^5$
that we will see presently.

\subsec{$SL(2,\Z)$ Invariance And Discrete Torsion}

The first important point about the Type IIB orientifold threeplane
is that there exists a supersymmetric orientifold threeplane that is invariant
under the $SL(2,\Z)$ $S$-duality symmetry group of Type IIB superstrings.
(This is implicit in the use \ref\kut{S. Elitzur, A. Giveon,
D. Kutasov, and D. Tsabar, ``Branes, Orientifolds, And Chiral
Gauge Theories,'' hep-th/9801020.} of this orientifold to
explain Montonen-Olive duality for orthogonal and symplectic gauge groups.)
To see this, recall first that the expectation values of the Type IIB dilaton
and axion fields  break $SL(2,\Z)$ down to a finite subgroup that generically
is generated by the element $-1$ of the center of 
$ SL(2,\Z)$.  This group element, which we will call $w$, is of order
two as an element of  $SL(2,\Z)$.  But on spinors it generates a transformation
of order four, not of order two.  In fact, one has $w^2=(-1)^F$, not
$w^2=1$.   This can be seen by inspection of the low energy Type IIB
supergravity.  If $Q_{\alpha\,L}$ and $Q_{\alpha\,R}$ are the left and
right-moving supercharges of the theory
(here $\alpha=1,2,\dots,16$ is a positive chirality spinor index of $SO(1,9)$;
the spinor index will sometimes be suppressed),
then $w$ acts by $w Q_{L}=Q_{R}$,   $w Q_{R}=
-Q_{L}$.  Or more succinctly, if we combine
$Q_{L}$ and $Q_{R}$ to a doublet $Q_{i}$,
$i=1,2$, we get $wQ_{i}=\epsilon_{ij}Q_{j}$; here 
$\epsilon_{ij}=-\epsilon_{ji}$ is
the antisymmetric tensor with $\epsilon_{12}=1$.

Now under an orientifolding operation that creates an orientifold
threeplane, the supersymmetries transform by
\eqn\hucx{M:Q_{\alpha i}\to (\Gamma^0\Gamma^1\Gamma^2\Gamma^3)_{\alpha\beta}
m_{ij}Q_{\beta j}}
with some  matrix $m_{ij}$ that
must map supersymmetries coming from left-movers to those coming from
right-movers.  So in the basis $Q_{ L},\,Q_{R}$, $m$
looks like
\eqn\oppo{m=\left(\matrix{0 & a \cr b & 0 \cr}\right)}
for some $a,b$.  The supersymmetry algebra obeyed by the $Q$'s implies
that $a$ and $b$ are each $\pm 1$.
  For $M$ to leave invariant some supersymmetry,
we must pick $m$ such that $M^2=1$; if instead $M^2=-1$ all supersymmetry
is broken.  As $(\Gamma^0\Gamma^1\Gamma^2\Gamma^3)^2
=-1$, this forces $m^2=-1$, so that $a=-b=\pm 1$.  Depending on the sign,
either $m$ equals the $SL(2,Z)$-invariant matrix $w$ described
in the last paragraph, or $m=w^3$.    
(The same argument shows that an orientifold
five-plane, for example, cannot be $SL(2,\Z)$-invariant; in that case,
one needs $m^2=1$, making it impossible for $m$ to commute with $w$.)

By using this $SL(2,\Z)$-invariant
orientifolding operation in the presence of threebranes, 
we can get an $SL(2,\Z)$-invariant
configuration of threebranes on $\R^4\times \R^6/\Z_2$, and hence
(after taking the near-horizon geometry) an $SL(2,\Z)$-invariant
compactification on $AdS_5\times \S^5/\Z_2$.  This is, however, not
the only possible model on $AdS_5\times \S^5/\Z_2$.  It is possible
to make additional models by turning on ``discrete torsion.''

To understand the possibilities, we must first understand how
the two two-form fields of the $SL(2,\Z)$ theory -- the Neveu-Schwarz
$B$ field $B_{NS}$ and the Ramond-Ramond  $B$ field $B_{RR}$ -- transform
under the orientifolding operation.  First of all, because the orientifolding
exchanges left- and right-movers on the string world-sheet, it reverses
the world-sheet orientation.  Hence $B_{NS}$ does not transform as an
ordinary two-form under orientifolding; the $\Z_2$ action multiplies
$B_{NS}$ by an extra minus sign. We can describe this by saying
that $B_{NS}$ is a twisted two-form: it is a section of $\Omega^2\otimes
\epsilon$, where $\Omega^2$ is the bundle of ordinary two-forms,
and  $\epsilon$ will denote the unorientable
real line bundle over
 $\RP^5$. Because the orientifolding
is $SL(2,\Z)$-invariant, $B_{RR}$, which is related to $B_{NS}$ by
$SL(2,\Z)$, likewise is a twisted two-form.

In fact, since $m$ acts on bosons as 
the element $-1\in SL(2,\Z)$,\foot{Since $w^2=(-1)^F$ acts trivially
on bosons, $w$ and $w^3$ both act on bosons as $-1\in SL(2,\Z)$.} it reverses
the sign of all string and fivebrane charges.  So both onebrane and fivebrane
orientations are reversed in going around a non-trivial loop in $\RP^5$.
This fact  will play an important role in the present paper.  

\bigskip\noindent{\it Topology Of The $B$-Field}

Let us temporarily consider a Type IIB string theory that has
{\it not} been orientifolded.
In such a theory, the gauge-invariant field strength of a two-form field $B$ is
a three-form $H=dB$.  Being closed, it determines a cohomology
class $[H]$ that takes values in $H^3(M,\R)$, where $M$ is the spacetime
manifold.  When discrete torsion is taken into account, the cohomology 
class that measures the topology of the $B$-field is actually an element
of $H^3(M,\Z)$.  We denote it in general as $[H]$ and call it the
characteristic class of the $B$-field.

Now return to the case of an orientifold.  In this case,
$B$ is a twisted two-form.
The gauge-invariant field strength is still $H=dB$, but now $H$ is
a twisted three-form (a section of $\Omega^3\otimes \epsilon$, with
$\Omega^3$ the bundle of ordinary three-forms).  
$H$ is still closed; it determines
a  cohomology class $[H]$ that now takes values not in $H^3(M,\R)$,
but in $H^3(M,\tilde \R)$, where $\tilde \R$ is the constant sheaf
$\R$ twisted by $\epsilon$ (it is the sheaf of locally constant sections
of $\epsilon$).  When discrete torsion is included, the topological
type of the $B$-field is measured not by an element of $H^3(M,\Z)$,
as in the previous paragraph, but by an element $[H]$ of $H^3(M,\tilde \Z)$,
where $\tilde \Z$ is a twisted sheaf of integers. Like $\tilde \R$,
$\tilde \Z$ is built using the
same $\pm 1$-valued transition functions
used in defining the real line bundle $\epsilon$; thus concretely,
as one goes around a noncontractible loop in $\RP^5$, a section of $\tilde \Z$
(or $\tilde \R$) comes back to itself with a reversal of sign.

\def\TZ{{\bf{\tilde Z}}}
For many subsequent applications, we will need to know the
homology and cohomology of $\RP^5$ with ordinary and twisted coefficients.
A basic fact here is that an $\RP^i$ subspace of $\RP^5$,
defined by a linear embedding $(x_1,x_2,\dots,x_{i+1})\to (x_1,x_2,\dots,
x_{i+1},0,\dots,0)$, is orientable or unorientable depending on whether $i$
is odd or even.  For odd $i$, the embedded $\RP^i$ determines an element
of $H^i(\RP^5,\Z)$, and for even $i$ it defines an element of $H^i(\RP^5,
\TZ)$.   For $1\leq i\leq 4$, these subspaces define two-torsion elements
that generate the respective homology groups.  The non-zero homology
groups are thus
\eqn\nobby{\eqalign{H_1(\RP^5,\Z)&=H_3(\RP^5,\Z)=\Z_2  ,\cr
       H_2(\RP^5,\TZ)&=H_4(\RP^5,\TZ)=\Z_2,\cr}}
along with $H_0(\RP^5,\Z)=H_5(\RP^5,\Z)=\Z, H_0(\RP^5,\TZ)=\Z_2$.

We will also have some use for the cohomology groups.  As $\RP^5$
is orientable, Poincar\'e duality tells us that
\eqn\hcb{\eqalign{H_i(\RP^5,\Z)& = H^{5-i}(\RP^5,\Z) \cr
                  H_i(\RP^5,\TZ) & = H^{5-i}(\RP^5,\TZ).\cr}}
Hence we have, in particular,
\eqn\mucu{H^3(\RP^5,\TZ)=\Z_2.}  
                   
\bigskip\noindent{\it The Four Models}

We can now classify the possible models.  The discrete torsion for
either $B$ field is classified by an element of $H^3(AdS_5\times \RP^5,\tilde
\Z)$ which (because $AdS_5$ is contractible) is the same as
$H^3(\RP^5,\TZ)=\Z_2$.  Hence, for either of the two $B$ fields, there
is precisely one possible non-trivial choice of discrete torsion.

One can describe explicitly what the choice of discrete                               
torsion means.   In general, including a $B$ field means that the
path integral for a string world-sheet $\Sigma$ (an elementary string
or $D$-string depending on whether we are considering $B_{NS}$ or $B_{RR}$)
receives an extra factor 
\eqn\ucuc{\exp\left(i\int_\Sigma B\right).}
In our problem, 
since the characteristic class of the $B$ field is a torsion
element, it can be represented by a $B$ field for which the curvature
$H=dB$ is zero.   This is the choice that we want to make
in order to get an $AdS_5\times \S^5$ compactification with
spacetime supersymmetry.  With this choice, the phase factor in \ucuc\ depends
only on the homology class
represented by $\Sigma$.  Since the orientation of
$\Sigma$ is supposed to change sign in going around a noncontractible
loop in $\RP^5$, the homology class of $\Sigma$ is an element of $H_2(\RP^5,
\tilde \Z)=\Z_2$.  The nontrivial element can be represented by $\Sigma=
\RP^2$, with a map to $\RP^5$ of the topological type given by a linear
embedding as in eqn.  \hunflo.  A zero-curvature $B$-field that is trivial
for such an $\RP^2$ would be completely trivial.  If non-trivial for
such an $\RP^2$, the $B$ field must multiply the world-sheet path
integral precisely by $-1$.   (The reason for this is that the homology
class of the embedded $\RP^2$ is a two-torsion element -- generating
$H_2(\RP^5,\tilde \Z)=\Z_2$
 -- and so must be weighted by a factor of square $+1$.)
This leads to a complete description of the role of the $B$-field,
as follows.  Any closed Riemann surface $\Sigma$ is the connected sum of
an oriented surface with $s$ copies of $\RP^2$, for some $s$.  
If $\Sigma$ is mapped to $AdS_5\times \RP^5$ by a map $\Phi$ that
obeys $\Phi^*(x)=w_1(\Sigma)$, it represents the same element of
$H_2(\RP^5,\TZ)$ as $s$ disjoint, linearly embedded $\RP^2$'s.
So the discrete torsion multiplies the
path integral for such a $\Sigma$ by a factor of $(-1)^s$.

We can now see that there are four possible models, depending on two
independent choices: the discrete torsion both for $B_{NS}$ and       
for $B_{RR}$  has two possible values, zero or non-zero.
Let us call the two types of discrete torsion $\theta_{NS}$ and $\theta_{RR}$.
The action of $SL(2,\Z)$ on the four models is easy to identify,
given what we know about $B_{NS}$ and $B_{RR}$.  The two $B$-fields
transform in the two-dimensional representation of $SL(2,\Z)$, so
$\theta_{NS}$ and $\theta_{RR}$ transform in that two-dimensional
representation, reduced mod    2.  So the model
with $(\theta_{NS},\theta_{RR})=(0,0)$ is $SL(2,\Z)$-invariant,
and the other three models are permuted by $SL(2,\Z)$
like the three non-zero elements
of $\half\Lambda/\Lambda$, with $\Lambda$ a lattice acted on by $SL(2,\Z)$
in the natural way.

Now let us determine the gauge groups of the four models.
A priori, we might have $SO(N)$ or $Sp(N/2)$ for some $N$.
The easiest model to identify is the $(0,0)$ model.  Since the $AdS_5
\times \S^5$ compactification without discrete torsion is 
$SL(2,\Z)$-invariant, the corresponding four-dimensional gauge theory
has Montonen-Olive self-duality.  The $\N=4$ theory with $SO(2k)$
gauge group has Montonen-Olive self-duality, while the $SO(2k+1)$ and
$Sp(k)$ theories are exchanged under duality.  So the model without
discrete torsion is an $SO(2k)$ gauge theory.

Now let us analyze the other models.  Turning on $\theta_{NS}\not=0$
multiplies the $\RP^2$ contribution to the worldsheet path integral of
fundamental strings by a factor of $-1$.  This has the effect
of exchanging symplectic and orthogonal gauge groups.  We already
know that one $\theta_{NS}=0$ model (the $(0,0)$ model) has orthogonal
gauge group.  So the
models with symplectic gauge group are the models
with $\theta_{NS}\not=0$; they are in other words the
$(\theta_{NS},\theta_{RR})=(1/2,0)$ and $(1/2,1/2)$ models.
Likewise the models with orthogonal gauge groups are $(0,0)$ and $(0,1/2)$.
We have already determined that the $(0,0)$ model has gauge group
$SO(2k)$ for some integer $k$;  the $(0,1/2)$ model, being related
by $SL(2,\Z)$ duality to the $(1/2,0)$ $Sp(k)$ model, has instead gauge
group $SO(2k+1)$.  The whole picture is portrayed in figure 3.
Notice that in this analysis we recover the correspondence between
$Sp(k)$ and $SO(2k+1)$ models and non-zero points in $\half\Lambda/\Lambda$
that was found in studies of their Coulomb branches after soft
breaking to $\N=2$ supersymmetry
 \nref\uranga{A. Uranga, ``Towards Mass Deformed
${\cal N}=4$ $SO(N)$ and $Sp(k)$ Gauge Theories From Brane Configurations,''
hep-th/9803054.}
\nref\yok{T. Yokono, ``Orientifold 4-Plane In Brane Configurations And
$\N=4$ $USp(2N_c)$ and $SO(N_c)$ THeory,'' hep-th/9803197.}
\nref\hoker{E. D'Hoker and D. H. Phong, ``Calogero-Moser Lax Pairs
With Spectral Parameter For General Lie Algebras,'' hep-th/9804124,
``Spectral Curves For Super-Yang-Mills With Adjoint Hypermultiplet
For General Lie Algebras,'' hep-th/9804126.}
\refs{\uranga - \hoker}.

\bigskip
\centerline{\vbox{\hsize=4in\tenpoint
\centerline{\psfig{figure=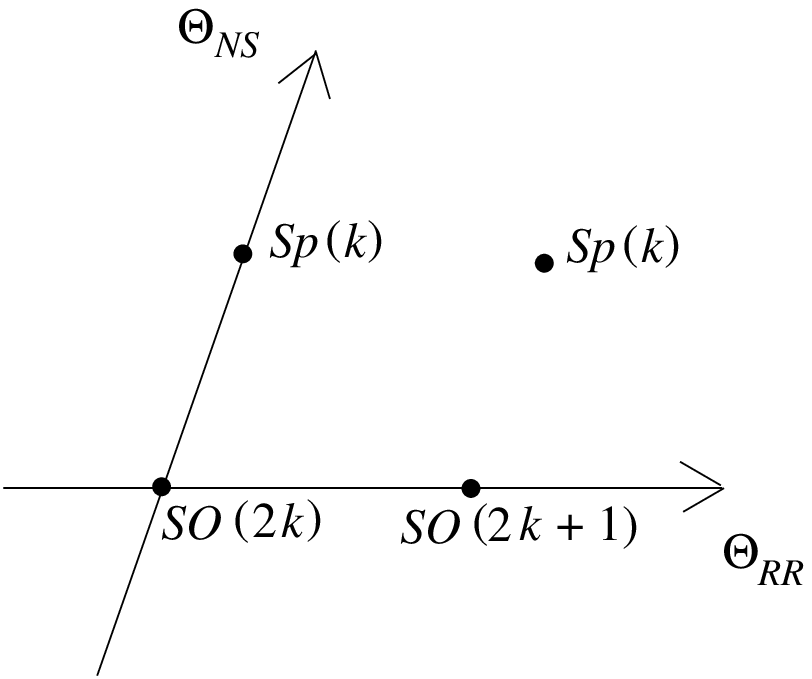}}
\vglue.4in
Fig. 3.  Sketched here are the four possible models, with zero
and non-zero values of the discrete theta angles $\theta_{NS}$ and 
$\theta_{RR}$.  Each model is labeled by the corresponding gauge group.
The coordinate axes have been slanted, because the four models correspond
to the half-lattice points of an arbitrary rank two lattice in the
complex plane.}}


\subsec{Possibilities Of Brane Wrapping}

The rest of this paper will be concerned primarily with interpreting
wrapped branes of various kinds in the $SO(N)$ and $Sp(N/2)$ gauge theories.
First we make a crude classification that is independent of the discrete
torsion, and then we consider an important refinement.

At first sight, it might appear that ten-dimensional strings 
(either elementary strings or $D$-strings) can be either wrapped
on a one-cycle in $\RP^5$ or unwrapped, giving either a zero-brane or
a one-brane on $AdS_5$.  However, the wrapping modes that would
give zero-branes are actually not allowed.  Such a wrapping
mode would correspond to a worldsheet of the form
$S\times \R$, with $S$ a circle mapped to
a noncontractible loop in $\RP^5$ and $\R$ the time direction.
As $S\times \R$ is orientable, such a world-sheet
does not obey the condition that the orientation should be reversed
in going around a noncontractible loop; and hence such wrapping modes
are not present.

A more formal way to reach this conclusion is to state that since the
string orientation is supposed to be reversed in going around the 
noncontractible loop, the strings are classified not by $H_1(\RP^5,\Z)$
but by $H_1(\RP^5,\TZ)$.  As this group vanishes, there is no nontrivial
topological class of strings.

Now we consider the wrapping of threebranes.  Since the threebrane charge
is invariant under the orientifolding operation, the threebrane orientation
is invariant in going around a loop in $\RP^5$.  Hence wrapping modes
of the threebrane are classified by the ordinary (untwisted)
homology of $\RP^5$.  To get from the threebrane an $i$-brane
on $AdS_5$, we must wrap it on a $3-i$-cycle in $\RP^5 $; these
are classified by $H_{3-i}(\RP^5,\Z)$.  Looking back to \nobby,
we see that there are three possibilities:

{\it (i)} A ten-dimensional threebrane that is not wrapped at all
remains a threebrane in $AdS_5$.

{\it (ii)} The threebrane can be wrapped on a one-cycle, classified
by $H_1(\RP^5,\Z)=\Z_2$, to give a two-brane on $AdS_5$.

{\it (iii)} The threebrane can be wrapped on a three-cycle, classified
by $H_3(\RP^5,\Z)=\Z_2$, to give a particle on $AdS_5$.

Now we move on to fivebranes (for the moment $NS$ and $D$ fivebranes
can be treated alike).  Since fivebranes are dual to onebranes,
the fivebrane charge is, like the onebrane charge, odd under going
around a noncontractible loop in $\RP^5$.  So wrapping modes of fivebranes
are classified by twisted homology.  We cannot consider a completely
unwrapped fivebrane, since   the six-dimensional world-volume
of  a fivebrane does not really fit into
$AdS_5$ (which is only five-dimensional).  
So the possibilities are as follows:

{\it(i)$'$} The fivebrane can be wrapped on a two-cycle, classified
by $H_2(\RP^5,\TZ)=\Z_2$, to give a threebrane in $AdS_5$.

{\it (ii)$'$} The fivebrane can be wrapped on a four-cycle, classified
by $H_4(\RP^5,\TZ)=\Z_2$, to give a string in $AdS_5$.

The interpretation of the wrapping modes just described will be the
main focus of the remainder of the present paper.

\bigskip\noindent{\it A Topological Restriction}

The branes just described are subject to an important topological
restriction.
A Dirichlet fivebrane can be wrapped on an $\RP^4\subset \RP^5$, to make
a string, only if  $\theta_{NS}$ vanishes; likewise, an NS fivebrane
can be so wrapped only if   $\theta_{RR}$ vanishes.
And a threebrane can be wrapped on an $\RP^3\subset \RP^5$,
to make a particle, only if $\theta_{NS}=\theta_{RR}=0$.

To explain these restrictions, we begin with  the case of the
NS fivebrane and the vanishing of $\theta_{NS}$.  We recall
that the meaning of having $\theta_{NS}\not=0$ is that, for
a linearly embedded $\RP^2\subset \RP^5$, one has
\eqn\juzz{\exp\left(i\int_{\RP^2}B_{NS}\right)=-1.}
What is written on the left hand side is in fact slightly oversimplified.
The $B_{NS}$ field is not really well-defined as a two-form;
it is subject to the gauge transformations $B_{NS}\to B_{NS}+d\lambda$
with $\lambda$ a twisted one-form.  In general, $B_{NS}$ can only
be defined as a twisted two-form locally, after taking a cover of spacetime
by small open sets; the twisted
two-forms on different open sets are related by
compatible gauge transformations in the overlaps of the open sets.
The situation is somewhat analogous to that of a connection $A$ on
a $U(1)$ bundle; for any closed circle $S$ in spacetime, one has a holonomy,
usually written by physicists as
\eqn\vuzz{\exp\left(i\int_SA\right)}
even though in fact $A$ is only defined locally as a one-form.

In the presence of an NS fivebrane, the situation is changed as follows.
There is a $U(1)$ gauge field $a$ on the fivebrane world-volume.  If
$f$ (defined locally as $da$) is the  curvature of $a$,
then the two-form $B'=B_{NS}-f$ is gauge-invariant,\foot{Under
$B_{NS}\to B_{NS}+d\lambda$, $a$ transforms as $a\to a+\lambda$.}
 and is hence well-defined
as an ordinary, global, twisted two-form.   
$B'$ is closed, since $B_{NS}$ is closed and $f$ is also closed by virtue
of the Bianchi identity.

Now, suppose
that the fivebrane is wrapped on an $\RP^4$, and deform the $\RP^2$
in \juzz\ to be a subspace of this $\RP^4$.  Because of the Dirac
quantization condition on the flux of $f$, $\exp(i\int_{\RP^2}f)=1$
(note that $f$ is a twisted two-form and so can be integrated over
the unorientable manifold $\RP^2$).  So we have
\eqn\buzz{\exp\left(i\int_{\RP^2}B_{NS}\right)=
\exp\left(i\int_{\RP^2}B'\right).}

However, for any closed globally defined twisted two-form $B'$, one has
\eqn\cuzz{\int_{\RP^2}B'=0.}
In fact, this integral is determined by the cohomology class of
$B'$ in $H^2(\RP^4,\TZ)$, but this group vanishes. 

So \buzz\ implies that 
\eqn\ofuzz{\exp\left(i\int_{\RP^2}B_{NS}\right) = 1}
or in other words that the Dirichlet fivebrane can be wrapped on
an $\RP^4\subset\RP^5$, to make a string in $AdS_5$,
 only if $\theta_{NS}$ vanishes.
Precisely the same argument (considering the path integral in a sector
with a $D$-string worldvolume wrapped on an $\RP^2\subset \RP^4$)
shows that the NS  fivebrane can be wrapped
on such an $\RP^4$, to make a string, only if $\theta_{RR}$ vanishes.
The same argument (considering wrapping on $\RP^2\subset\RP^3$ of
a fundamental or $D$-string worldvolume) shows that a threebrane
can be wrapped on $\RP^3$ to make a particle on $AdS_5$ only
if $\theta_{NS}=\theta_{RR}=0$.  Here one uses $H^2(\RP^3,\TZ)=0$ to
show vanishing of \cuzz.

There is, however, no such restriction on wrapping of fivebranes
on $\RP^2\subset\RP^5$ to make threebranes.  Since $H^2(\RP^2,\TZ)=\Z$,
\cuzz\ need not vanish.  There is no restriction on wrapping of a threebrane
on $\RP^1$, since in this case $\RP^2$ cannot even be deformed into the
threebrane.

\bigskip\noindent{\it General Formulation And Curvature Correction}

The above has been formulated in a rather {\it ad hoc} fashion.
The basic idea is really the following.
In the situation considered above, along the brane world-volume $Y$,
we asserted the existence of a gauge-invariant field $B'=B-da$ that is
gauge-equivalent to $B$.
Existence of such a field means that the characteristic class $[H]$
of the $B$-field, when restricted to $Y$, vanishes as an element of
$H^3(Y,\TZ)$.  For instance, as a differential form the field strength
of $B$ is $H=dB'$, and this is certainly zero in cohomology.
However, existence of a globally-defined $B'$ field that is gauge-equivalent
to $B$ means that $[H]$ vanishes, when restricted to $Y$, as an integral
class, not just in real cohomology.  The analog of this assertion for
one-form fields is perhaps more familiar.  If a complex line bundle
${\cal L}$ has a connection that can be represented by a globally-defined
one-form $A$, then ${\cal L}$ is topologically trivial and
$c_1({\cal L})$ vanishes in integral cohomology,
not just in real cohomology. 

It was essential in deriving the restrictions on brane wrapping that
the restriction of $[H]$ to $Y$ vanishes in integral (and not just real)
cohomology, since in the situation we considered, $[H]$ is anyway
a torsion class.  So let us look at this point closely.  We write $i$
for the inclusion of $Y$ in spacetime, and $i^*([H])$ for the restriction
of $[H]$ to $Y$.  The gauge invariance of $B'=B-da$ with respect to local
gauge transformations is certainly well understood in perturbative
string theory.
However, we should ask about invariance under global gauge transformations.
Failure of such invariance would be a sort of ``global anomaly''
in the gauge-invariance of $B'$, depending on the topology of $Y$,
and might mean that $i^*([H])$ is not zero but is a torsion class determined
by $Y$.   ($i^*([H])$ would have to be a torsion class, since the formula
$B'=B-da$ is certainly well enough understood to show vanishing of
$i^*([H])$ at the level of  differential forms.)
If from the differential
topology of $Y$, we can build in a natural way
 a torsion class $W\in H^3(Y,\TZ)$,
then we should consider the possibility that the correct global
restriction is not that $i^*([H])=0$  but rather
that
\eqn\kiloo{i^*([H])=W.}

A possible $W$ can be built as follows.  Begin with the
second Stieffel-Whitney class $w_2(Y)\in H^2(Y,\Z_2)$.  Consider the
exact sequence
\eqn\iloo{0\to \TZ\underarrow{2} \TZ\to \Z_2\to 0,}
where the first map is multiplication by 2 and the second is reduction
modulo 2.  The ``connecting homomorphism'' in the long exact sequence
of cohomology groups that is derived from \iloo\ maps $w_2$ to
an element $W\in H^3(Y,\TZ)$.  We can thus contemplate that the correct
global condition is  $i^*([H])=W$ rather than $i^*([H])=0$.  
Whether such a correction is needed to cancel worldsheet
global anomalies is under investigation \ref\frw{D. Freed and E. Witten, 
to appear.}.  Such a correction would be more or less analogous to
the gravitational shift in the quantization law of the four-form
field strength of $M$-theory \ref\uwitten{E. Witten, ``On Flux Quantization
IN $M$ Theory And The Effective Action,'' J. Geom. Phys. {\bf 22}
(1997) 1.}.  We will describe in section 4 a further contribution to 
\kiloo\ that arises when additional branes of lower dimension end on
the threebrane or fivebrane.

For most of this paper, it will not matter whether the correct condition
on $H$ contains the correction term $W$ or not.  The reason for this
is that we will be considering branes wrapped on $\RP^i$ with $i\leq 4$
or $\S^5$, and these spaces all have $W=0$.  However, when we come
to the question of stability of the baryon vertex for orthogonal
and symplectic gauge groups, we will reach a situation in which the
correction term is apparently needed to make $AdS$ string theory match
correctly with gauge theory predictions.

\newsec{Gauge Theory And Branes On $\RP^5$}

In this section, we will consider the various types of branes
in $AdS_5\times \RP^5$, and interpret them in orthogonal
and symplectic gauge theory.  We consider in turn particles,
strings, domain walls, the baryon vertex, and instantons.

\subsec{The Pfaffian}

First, we consider particles 
obtained by wrapping a threebrane
on $\RP^3\subset \RP^5$.  Since this $\RP^3$ represents a generator
of $H^3(\RP^5,\Z)=\Z_2$, the particle obtained this way is stable,
but the number of such particles is conserved only modulo two;
a pair of such particles can annihilate.

Now we apply the topological restriction found in section 3.3.
A threebrane can wrap on $\RP^3$ only if both $B$-fields are
topologically trivial, that is only if $\theta_{NS}=\theta_{RR}=0$.
In view of the classification of the four models that was summarized in
figure 3, this means that
the gauge group is $SO(2k)$ for some $k$, and not $SO(2k+1)$ or $Sp(k)$.
$k$ is of course the number of five-form flux quanta on $\RP^5$:
\eqn\hxv{\int_{\RP^5}{G_5\over 2\pi}=k.}
On the double cover $\S^5$, the number of quanta is $N=2k$.

So we have found particles, conserved modulo two, that exist precisely
for $SO(2k)$ gauge group, and not for the other symplectic or orthogonal
groups.  There are in gauge theory such states -- the Pfaffians,
described in the introduction (created by acting on the vacuum with
a gauge-invariant operator of the form ${\rm Pf} (\Phi)$, with $\Phi$
a field in the adjoint representation and ${\rm Pf}$ the Pfaffian).  
One naturally suspects that the
threebrane wrapped on $\RP^3$ should be identified with the Pfaffian
particle of $SO(2k)$ gauge theory.

The number of Pfaffian objects is of course conserved modulo two
for a simple group-theoretical reason.  ${\cal N}=4$ super Yang-Mills theory
with gauge group $SO(N)$ actually has $O(N)$ symmetry; the quotient
$O(N)/SO(N)\cong \Z_2$ behaves as a global symmetry group.  If
$\tau$ is the generator of this $\Z_2$, then the Pfaffian is odd under
$\tau$ and cannot decay to objects that are even under $\tau$.
Note that any gauge-invariant state constructed with less than $N/2$
elementary quanta is even.  Since the global symmetry group is $\Z_2$,
Pfaffians can annihilate in pairs, just like wrapped threebranes.

As further evidence for  identifying Pfaffians with wrapped threebranes, 
we will determine the
quantum numbers of the low-lying states on the two sides under the
$R$-symmetry group of the theory.
First we begin in the $AdS$ description, where the $R$-symmetry group
is a cover of the symmetry group of $\RP^5$.  The manifold $\RP^5$ 
(identified as the sphere $\sum_{i=1}^6x_i^2=1$ modulo the $\Z_2$
symmetry $x_i\to -x_i$)
has a symmetry group $G_0=SO(6)/\Z_2$.  
A given $\RP^3$ subspace, say $x_5=x_6=0$, is invariant under
$H_0=(SO(4)\times SO(2))/\Z_2$ (where $SO(4)$ acts on the first four
coordinates and $SO(2)$ on the last two).    The space of such
embeddings is thus the homogeneous space $G_0/H_0$, which is the
same as $G/H$, with $G=SO(6)$ and $H=SO(4)\times SO(2)$.

To find the low-lying threebrane quantum states, we must ``quantize the
collective coordinates,'' and analyze the quantum mechanics on $G/H$.
The quantum wave states are not ordinary functions
on $G/H$, but sections of a line bundle of degree $k$.  The line
bundle appears because the threebrane is electrically charged with respect
to the five-form field strength $G_5$, and is of degree $k$ (that is,
it is the $k^{th}$ power of the most basic line bundle) because the number
of flux quanta on $\RP^5$ is $k$.  The Hamiltonian acting on such
functions is a $G$-invariant Laplacian.  In our present case, $G/H$ is
a symmetric space, and there is only one invariant Laplacian.

We can identify $G/H$ as the $G$ manifold, with $H$ acting on the right
and $G$ on the left.  Functions on $G/H$ are simply $H$-invariant functions on
$G$, that is functions $\psi(g)$ with $\psi(g)=\psi(gh)$ for all $h\in H$.
A section of a line bundle on $G/H$ is  a function on the 
$G$ manifold that obeys
\eqn\nurky{\psi(gh)=\psi(g)r(h).}
Here $h\to r(h)$ is a homomorphism of $H$ to $U(1)$; the choice of
homomorphism determines the line bundle.  
In our case, with $H=SO(4)\times SO(2)$, 
the most general homomorphism of $H$ to $U(1)$ is the product of the
``charge $s$'' representation of $SO(2)$ (for some $s$) and the trivial
representation of $SO(4)$.  We want to set $s=k$, to get a line bundle of
degree $k$, so for us
$\psi $ should be $SO(4)$-invariant
and should transform with charge $k$ under $SO(2)$.

\def\6{\bf 6}
\def\4{\bf 4}
If we identify $G=SO(6)$ as the group of $6\times 6$ orthogonal matrices
$g^i{}_j,\,i,j=1,\dots, 6$, then functions on the $G$ manifold can be expanded 
as polynomials in the matrix elements of $G$.  The matrix elements $g^i{}_j$ 
themselves transform as $(\6,\6)$ under $SO(6)\times SO(6)$, and as
$(\6,\4)^0\oplus (\6,{\bf 1})^1\oplus(\6,{\bf 1})^{-1}$
under $SO(6)\times SO(4)\times SO(2)$; here the exponent is the $SO(2)$
charge.  To make sections of the desired line bundle, we want polynomials
in the $g^i{}_j$ of $SO(2)$ charge $k$.  To minimize the energy,
that is the eigenvalue of the Laplacian, we must select 
the polynomials of lowest
degree that have charge $k$.  These are simply
the  polynomials of degree $k$ in
the $(\6,{\bf 1})^1$.  Note that these polynomials transform in the
{\it traceless} symmetric product of $k$ copies of the $\6$; the trace
terms vanish because $g$ is an orthogonal matrix.  

The manifold $SO(6)/(SO(4)\times SO(2))$ is actually a homogeneous
Kahler manifold, and the line bundle just considered is ample.
\foot{To be more precise, $SO(6)/(SO(4)\times SO(2))$ is
the same as $SU(4)/(SU(2)\times SU(2)\times U(1))$ and is the Grassmannian
of complex two-planes in ${\bf C}^4$; it can also be described
as a quadric in ${\bf CP}^5$.  The line bundle described in the
last paragraph is the usual very ample line bundle over this Grassmannian or 
quadric.}
Hence the lowest-lying wavefunctions just found are holomorphic
sections of the line bundle.  They consequently give BPS states.  It is
therefore  possible to compare them to chiral operators in the boundary
conformal field theory.  Such operators can be constructed 
as in \ref\seiberg{N. Seiberg, ``Notes On Theories With 16
Supercharges,'' hep-th/9705117.} from the
scalar fields  of the $\N=4$ super Yang-Mills theory
on the boundary.  The scalars transform in the ${\bf 6}$ of the $SO(6)$
global symmetry and in the adjoint of the gauge group $SO(2k)$.
The Pfaffian of the scalars (that is, $\epsilon^{a_1a_2\dots a_{2k}}
\Phi^{i_1}_{a_1a_2}\Phi^{i_2}_{a_3a_4}\dots \Phi^{i_k}_{a_{2k-1}a_{2k}}$,
where the $i$'s are $SO(6)$ indices and the $a$'s are gauge indices)
transforms in the $k^{th}$ symmetric product of the ${\bf 6}$.  
The traceless part of the $k^{th}$ symmetric product can be shown
to consist of BPS operators by considering an $\N=1$ subalgebra of the
supersymmetry algebra.

Because of the BPS property, it is not really necessary to compare
the precise masses and dimensions of operators, but it is instructive
to work out the order of magnitude.  A threebrane wrapped on a volume
$V$ has a mass $m$ in $AdS$ units of order $V/\lambda$, with $\lambda$ the
string coupling constant.  According to \refs{\kleb,\witten}
the dimension of the corresponding conformal field theory operator
is of order $mR$, with $R$ the radius of curvature of $AdS$.  In the present
case, $V\sim R^3$ and $R\sim (\lambda k)^{1/4}$, so the dimension
is  $k$ in order of magnitude, and independent of $\lambda$.  
This agrees with conformal
field theory, where the Pfaffian operator has dimension precisely $k$.
The BPS property of course ensures that the coefficient of $k$
 also works out correctly.

\bigskip\noindent{\it String Ending On Threebrane}

\bigskip
\centerline{\vbox{\hsize=4in\tenpoint
\centerline{\psfig{figure=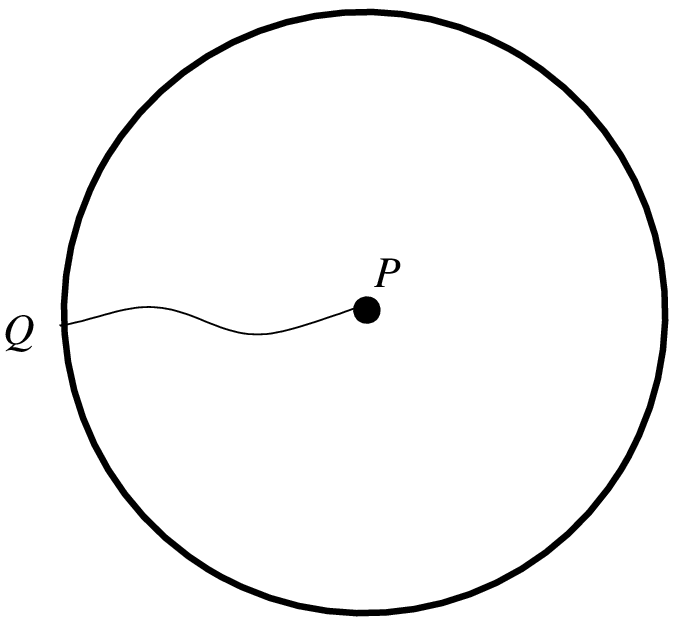}}
\vglue.4in
Fig. 4. For gauge group $SO(2k+1)$, a threebrane wrapped on
$\RP^3$ must have an odd number of strings ending on it.  Sketched
in the figure is a threebrane, at a point in $AdS$ space labeled $P$,
connected to the boundary by an elementary string that terminates
on a boundary point labeled $Q$. }}

So far we have considered wrapped threebranes without strings.
But a basic property of the threebrane is that a string can end on it.
 Let us consider a situation in which a single elementary Type IIB
string, as in figure 4, ends on a threebrane that is wrapped on $\RP^3$.
We suppose that the other end  of the string is connected to an elementary
quark (in the fundamental representation of the gauge group) on the
boundary of $AdS$ space.

Consider, in general, a Type IIB string ending on a $D$-brane.  Generally,
the end of the string is electrically charged with respect to the $U(1)$
gauge field $a$ living on the $D$-brane.  Let $f$ be the field strength
of $a$.  In the special case of a threebrane, we can make a duality
transformation to a dual $U(1)$ gauge field $\tilde a$ and field
strength $\tilde f=*f$.  The endpoint of the Type IIB string is
magnetically charged with respect to $\tilde f$:
\eqn\ovvo{d\tilde f=2\pi \delta_P,}
where $\delta_P$ is a delta function supported at the
endpoint of the string.  ($\delta_P$ is normalized so that its integral is
$1$; its coefficient is $2\pi$, since the string carries one flux quantum.)

We recall now that the field strength $H_{RR}$ of the Ramond-Ramond two-form 
$B_{RR}$ is in the absence of any branes
$H_{RR}=dB_{RR}$.  Along the brane, there is a gauge-invariant version
 $B'_{RR}=B_{RR}-\tilde f$ of the $B$-field.  Rewriting the definition
 of $H_{RR}$ in terms of $B'_{RR}$, we get
\eqn\impuu{H_{RR}=dB'_{RR}-2\pi\delta_P.}
There is no such magnetic correction for $B_{NS}$, whose field
strength remains
\eqn\bimpuu{H_{NS}=dB'_{NS}.}

A point  $P\in \RP^3$ generates $H_0(\RP^3,\TZ)=\Z_2$, and its
Poincar\'e dual $\delta_P$ generates $H^3(\RP^3,\TZ)$, which is likewise
$\Z_2$.  So the formulas \impuu, \bimpuu\ say that in the presence of the 
string, $H_{RR}$ is topologically non-trivial while $H_{NS}$ is trivial;
thus the wrapped threebrane, with a single string ending on it,
is possible only if $\theta_{RR}\not= 0$, $\theta_{NS}=0$.  This
configuration is in other words possible if and only if the gauge
group is $SO(2k+1)$ for some $k$.\foot{Note that a $B$-field on 
$\RP^3$ whose field strength $H$ has the delta function contribution
in \impuu\ (and which is trivial away from $P$) is topologically
equivalent to a $B$-field whose field strength vanishes and which
assigns the value $-1$ to a wrapped $\RP^2$.  
They are topologically equivalent
as there is in fact only one nonzero element of $H^3(\RP^3,\TZ)$.}

This has a natural interpretation in gauge theory.  From
a field $\Phi$ in the adjoint representation of $SO(2k+1)$, we cannot
make a gauge-invariant Pfaffian.  But we can couple $k$ copies of $\Phi$
to make a field $\epsilon^{ab_1b_2\dots  b_{2k}}\Phi_{b_1b_2}\dots
\Phi_{b_{2k-1}b_{2k}}$ in the fundamental representation of the gauge
group.  So in an $SO(2k+1)$ gauge theory
(such as the $\N=4$ supersymmetric theory) in which all elementary
fields are in the adjoint representation, an external quark in the
fundamental representation can be screened, by a combination of $k$ gluons.
In the $AdS$ description, this screening is described by connecting
the external quark to a wrapped threebrane by an elementary Type IIB string.

A more sophisticated way to describe \impuu\ is that the
trivialization of the $B$-field along the threebrane fails in the
presence of a string boundary.  If $C$ is the curve on the threebrane
worldvolume along which the string ends (thus $C=P\times {\bf R}$ in
a static situation, with $\R$ parametrized by time), then the condition
$i^*([H])=0$ discussed at the end of section 3 is corrected in the
presence of the string to
\eqn\recallit{i^*([H_{RR}])=[C],}
with $[C]$ the Poincar\'e dual to $C$.
In this equation,    $C$ is the boundary of the elementary strings that
end on the threebrane; there is a similar equation with $H_{RR}$ replaced
by $H_{NS}$ and $C$ replaced by the boundary of the $D$-strings ending
on the threebrane.

\bigskip\noindent{\it Particles and Operators}

A clarification should be added here.  Let us go back to the $SO(2k)$ case,
and reconsider the threebrane wrapped on $\RP^3$ with no strings attached.
Such a wrapped threebrane is a particle from the $AdS$
point of view, but does not quite have that interpretation in the boundary
theory because of the absence of a mass gap. Describe  $AdS_5$ 
by the metric $ds^2=f(x_0)(dx_0^2+\sum_{i=1}^4dx_i^2)$, where
$f(x_0)=1/x_0^2$, and the $x_i,\,
i>0$ should be
understood as the coordinates of the boundary theory which is
defined at $x_0=0$.  An $AdS$ particle of mass $m$, located at a given
$x_0$, has energy $m\sqrt{f(x_0)}$ from the point of view of the boundary
theory; because $f(x_0)\to 0 $ for $x_0\to \infty$, this can be arbitrarily
small.  Hence the wrapped threebranes considered above should be associated
in the boundary conformal field theory with Pfaffian operators, which
acting on the vacuum can create states of arbitrarily small energy.
Suppose that  as in various constructions in \otherwitten, one replaces $AdS_5$
by a  similar metric   with $f(x_0)$ bounded strictly above zero, and let
$f_0$ be the minimum value of $f$.
Then the boundary theory has a mass gap, and a particle of mass $m$ in
the interior theory gives rise to a particle of mass roughly
$mf_0$ in the boundary theory.  By this mechanism, in a suitable context,
one would describe actual Pfaffian particles in gauge theory on the boundary
in terms of wrapped branes in the interior.

\subsec{Strings}

We now consider strings in $AdS_5$ that arise by wrapping a
Dirichlet fivebrane on $\RP^4\subset \RP^5$.  We will call these
strings fat strings to avoid confusing them with elementary
strings and $D$-strings.

According to the criterion in section 3.3, fivebrane wrapping on $\RP^4$
is only possible if $\theta_{NS}=0$, that is
if the gauge group is orthogonal rather than symplectic.
Fat strings can annihilate in pairs, because they are classified
by $H_4(\RP^5,\TZ)=\Z_2$.  Their tension is proportional to the
Dirichlet fivebrane tension, 
and so is of order $1/\lambda$, that is, of order $N$.\foot{As 
in the last paragraph of section 4.1, this string tension
in $AdS$ space will become a string tension in the boundary theory only
if the boundary theory is perturbed to have $f_0>0$.  But even without
making such a perturbation, the factor of $N$ is observable in the boundary
theory.  For example, in the boundary conformal field theory, the energy
of a state with external spinor charges at specified locations on the boundary
is of order $N$, since it receives a contribution from the fat
strings connecting them in the interior of $AdS$ space.}

Strings with precisely these properties are expected in the boundary
conformal field theory.  They are strings associated with external
charges in the spinor representation of the gauge group.
This representation does not arise in the tensor
product of any number of copies of the $N$-dimensional or fundamental
representation of
$SO(N)$, so an external spinor charge is associated with a new kind of string
that cannot decay to the strings associated with charges in the fundamental
representation.
The new strings are conserved only modulo two, since the tensor product
of two spinors can be decomposed as a sum of tensor products of
the fundamental representation.  No such strings are expected for
symplectic gauge group, since the symplectic group has no ``new'' 
representations beyond what one finds in the tensor products of the fundamental
representation.  
It is natural for the string associated with an external spinor charge
to have tension of order $N$, since the highest weight vector of
the spinor representation (which is the vector $(1/2,1/2,\dots,1/2)$,
with $N/2$ entries) has length of order $\sqrt N$.
Indeed, in a current algebra description \ref\curwitten{E. Witten,
``Current Algebra, Baryons, and Quark Confinement,'' Nucl. Phys.
{\bf B223} (1983) 433.}, strings associated with external spinor
charges are seen as open string solitons (which can sometimes
be deformed to $D$-branes) and have tensions of order $N$.

All of these facts encourage the idea that the fat string of $AdS_5$
is related to spinor charges on the boundary.
For more such evidence,
we will now go back to the orientifold whose near-horizon geometry is
$AdS_5\times \RP^5$.  By studying the orientifold, we will also get some
clues that will enable us to more precisely match group theory with the wrapped
fivebrane.

\bigskip\noindent{\it Orientifold And Fivebrane}

We consider the familiar $\R^4\times (\R^6/\Z_2)$ orientifold.
For $\R^4$ we take coordinates $x^0,\dots,x^3$, and for $\R^6$ we take
coordinates $x^4,\dots, x^9$.  The $\Z_2$ acts by $x^i\to -x^i$, $i=4,\dots,9$.
There are $N$ threebranes at $x^4=\dots=x^9=0$.

The $\R^6/\Z_2$ factor is interpreted as follows in $AdS_5\times
\RP^5$.  The radial function $\rho=\sqrt{\sum_{i=4}^9x_i^2} $ of $\R^6/\Z_2$
becomes
one of the $AdS_5$ coordinates, the other four being $x^0,\dots,x^3$.
The angular directions in $\R^6/\Z_2$ are identified with $\RP^5$.

Now consider a fivebrane whose world-volume is specified by
$x^1=x^2=x^3=x^9=0$, with arbitrary values of $x^0$ and of $x^4,\dots,x^8$.
From the $AdS_5\times \RP^5$ point of view, such a fivebrane is wrapped
on an $\RP^4\subset \RP^5$ and looks like a fat string on $AdS_5$.
The $\RP^4$ in question is the subspace of $\RP^5$ with $x^9=0$.  
The fat string worldsheet in $AdS_5$ is parametrized by $\rho$ and $x^0$
and is at $x^1=x^2=x^3=0$.  

The $5-3$ strings connecting the fivebrane to the threebrane are, in their
ground state, fermions \green; we already exploited this fact
in section 2 in our discussion of the baryon vertex.  In the present
case, since the fivebrane and the threebranes actually meet at 
$x^1=\dots=x^9=0$, the ground state of the $5-3$ string has zero energy.
Because of the orientifolding, the $5-3$ and $3-5$ strings are actually
equivalent.  The ground states of these strings give, overall, $N$ fermion
zero modes in the fundamental representation of $SO(N)$. 
Upon quantizing these fermion zero modes, we learn that the ground
state of the system transforms in the spinor representation of $SO(N)$
(as we will discuss momentarily, both chiralities appear if $N$ is even).
We denote the fermion zero modes as $\psi_1,\dots,\psi_N$; they
generate a Clifford algebra.

The fivebrane, interpreted as a string in $AdS_5$, has an endpoint at $\rho=0$.
This endpoint lies on the boundary of $AdS_5$, so this is
 an example of a fat string  that ends at a boundary
point of $AdS_5$.  We have seen, in this particular example, that there
is a charge in the spinor representation of $SO(N)$ at the boundary point.
This gives strong confirmation for the proposal that 
in general strings of this
type can terminate at points on the boundary of $AdS_5$ at which there
are external spinor charges.

In the absence of orbifolding, there is a $U(1)$ gauge field on the
fivebrane worldvolume; the $5-3$ and $3-5$ strings have respectively
charge $1$ and $-1$.  Orbifolding reverses the sign of the $U(1)$,
exchanging the $5-3$ and $3-5$ strings.  The gauge symmetry on the
fivebrane world-volume is broken down to $\Z_2$.  The $5-3$ strings
are all odd under this $\Z_2$, and in particular that is true for 
the Clifford algebra generators $\psi_1,\dots,\psi_N$.  The symmetry
of the Clifford algebra under which the generators are all odd is, 
for even $N$, usually called chirality -- it assigns the value $+1$
or $-1$ to spinors that transform in the two different spin representations
of $SO(N)$.  For odd $N$, there is only one spin representation, but there
are two inequivalent representations of the Clifford algebra -- distinguished
by the sign of the product $\psi_1\psi_2\dots\psi_N$, which commutes with
the Clifford algebra -- and the operation $\psi_i\to -\psi_i$ exchanges
these two representations.  The physical interpretation of the appearance
of both representations of the Clifford algebra is somewhat unclear,
so in returning to $AdS_5$, I will consider only the case of even $N$.

\bigskip\noindent{\it Return To $AdS_5$}

We consider in $AdS_5$ a string, made by wrapping a fivebrane
on $\RP^4$, that connects two boundary points of $AdS_5$.  
There is an external spinor charge at each end of the string.
Since the unbroken $\Z_2$ along the string is a gauge symmetry,
the overall quantum state of the string should be $\Z_2$-invariant,
provided we include the charges at the ends of the string.  We take
this to mean that the product of the total $\Z_2$ charge of the string
with the chiralities of the spinor charges at the ends of the string
equals $+1$.

\bigskip
\centerline{\vbox{\hsize=4in\tenpoint
\centerline{\psfig{figure=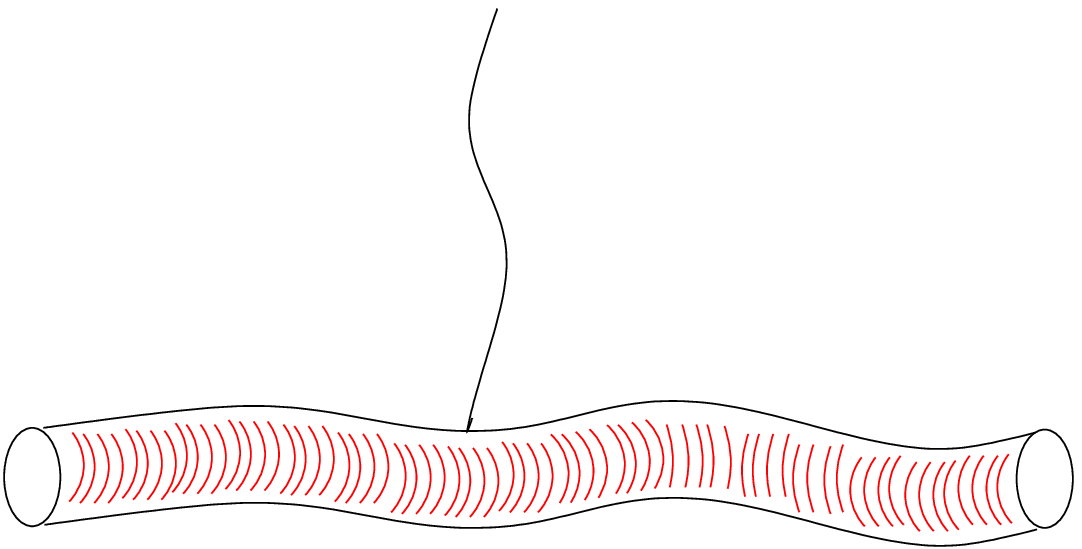}}
\vglue.4in
Fig. 5.  A ``fat string'' in $AdS$ space with a thin or elementary
string ending on it.  The fat string reverses chirality when absorbing
an elementary string.}}

Hence if, as in figure 5,
a fat string absorbs an elementary string, its chirality is
reversed.  In fact, the endpoint of an elementary string on a fivebrane
has charge $\pm 1$ under the $U(1)$ gauge field on the fivebrane -- and
so is odd under the unbroken $\Z_2$.  If, in joining on an elementary
string to the fivebrane, the $\Z_2$ quantum number to the left of the 
junction is kept fixed, then the chirality of the spinor charge on the
right must be reversed to preserve overall $\Z_2$ neutrality.  

This is in agreement with the following group-theoretical fact.
Let $S_+$ and $S_-$ denote the positive and negative chirality representations
of $SO(N)$ (for even $N$), and let $V$ be the vector  representation.
Then $V\otimes S_+$ contains $S_-$, but not $S_+$, and conversely
$V\otimes S_-$ contains $S_+$.  In other words, the chirality of a spinor
is reversed whenever a vector is absorbed, just as we find in analyzing
the fat strings.  

We now wish to reproduce one additional fact of group theory:
the tensor product $S_+\otimes S_+$ is a sum of representations of the
form $\wedge^s V$ (the $s^{th}$ antisymmetric tensor power of $V$) with
$s$ congruent to $N/2$ mod 2.  In other words, when two identical
fat strings annihilate, the number of elementary strings produced
should be $N/2$ mod 2.  We will see this in the following way.

\bigskip
\centerline{\vbox{\hsize=3in\tenpoint
\centerline{\psfig{figure=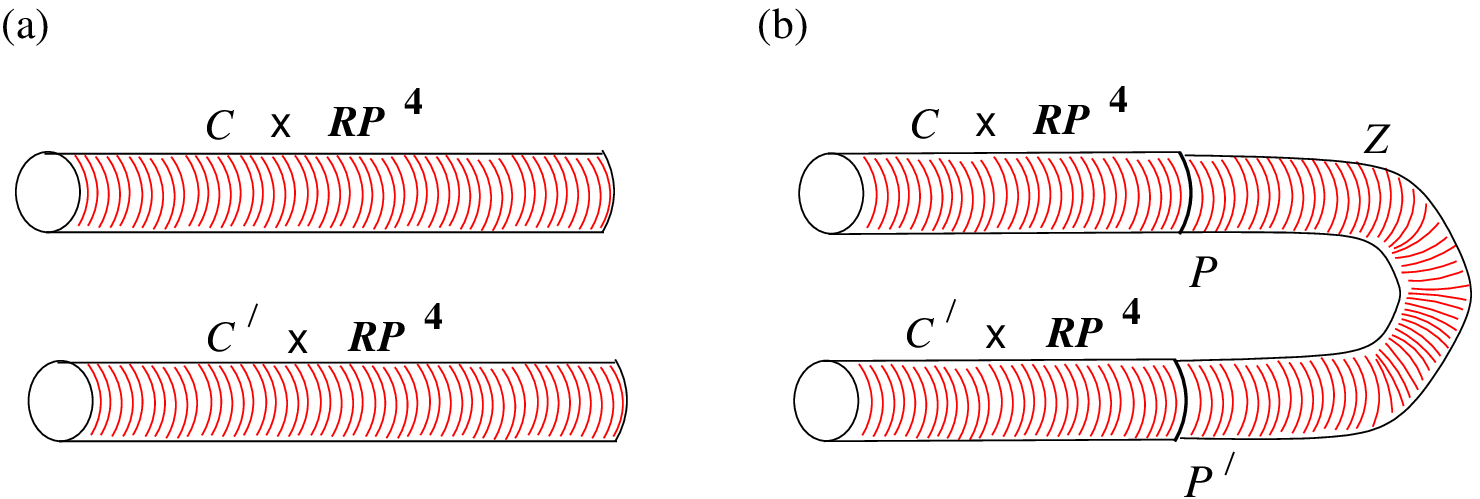}}
\vglue.4in
Fig. 6. Sketched in $(a)$ are two roughly parallel fat strings in $AdS$ space.
Sketched in $(b)$ is a process, described in the text, in which
the fat strings annihilate. }}

As a preliminary, recall that a fivebrane wrapped on $\RP^4$ is
unorientable.  For fivebranes in $AdS_5\times \RP^5$,
what  must be oriented is  not the tangent bundle
$TX$ of the fivebrane worldvolume $X$, but $TX\otimes \epsilon$,
where $\epsilon$ is the pullback to $X$ of the unorientable real line
bundle over $\RP^5$.  For any manifold $Z$ with tangent bundle $TZ$, we 
call an orientation of $TZ\otimes \epsilon$ a ``twisted orientation'' of $Z$.

Now, consider as in figure 6(a), two identical adjacent
 fivebranes, whose
worldvolumes, at given time, are of the form $C\times \RP^4$ and $C'\times
\RP^4$.  Here $C$ and $C'$ are two roughly parallel paths in $AdS_5$, and
for each fivebrane we use the same $\RP^4$.\foot{
Instead of thinking of the configuration in figure 6(a) as a
time-independent configuration with strings running from left to right,
one might alternatively want to think of the $AdS_5$ coordinate
that runs from left to right in this picture  as the ``time'' coordinate,
while an additional ``space'' coordinate that will be common to all
fivebranes is suppressed in the figure.  This way of looking
at the figure  agrees better with the terminology we will use
about ``annihilation'' of fat strings.}    
For example, if $\RP^5$ is obtained by projectivizing a copy of $\R^6$
with coordinates $x_1,\dots, x_6$, then $\RP^4$ can be defined as
the subspace 
\eqn\ikx{x_6=0.}
To ensure that the fivebranes
are identical, we want to give the ``same'' twisted orientation
to $C\times \RP^4$ and $C'\times \RP^4$, using the fact that $C$ and $C'$
are parallel and the two $\RP^4$'s are identical.

To describe the annihilation of the two fivebranes, we wish to suppose
that as in figure 6(b), $C$ and $C'$ are semi-infinite paths, terminating
at points $P$ and $P'$, respectively.
We pick a five-manifold $Z$ in spacetime whose boundary is the union
of $P\times \RP^4$ and $P'\times \RP^4$.  $Z$ should have a twisted
orientation which agrees on the boundary with those of $C\times \RP^4$
and $C'\times \RP^4$.  Then figure 6(b) describes the termination of 
the two semi-infinite fat branes upon arriving in the vicinity of $Z$.
We call this annihilation of the fat strings.

What can we pick for $Z$?  One is tempted to try  $Z=D\times \RP^4$
where $D$ is a path in $AdS_5$ from $P$ to $P'$, parametrized say by
an angle $\theta$, with $0\leq \theta\leq \pi$, and $\RP^4$ is still
the subspace \ikx\ of $\RP^5$.  With this choice of $Z$, however,
the twisted orientations in figure 6(b) would not match.  The problem
arises because as both $C$ and $C'$ are ``incoming,'' their
annihilation via $D$ involves a reversal of orientation.  The $\RP^4$
factor, being constant, has inevitably a constant twisted orientation,  
so consideration of this factor does not help.

To fix things, one must let the $\RP^4$ factor vary in such a way that
as one goes from $\theta=0$ to $\theta=\pi$, $\RP^4$ comes back with the
opposite twisted orientation, thus accounting for the orientation 
reversal that is involved in letting $C$ and $C'$ annihilate.
To make this happen, we replace \ikx\ with the
$\theta$-dependent condition
\eqn\jurrys{\cos\theta \,x_6+\sin\theta \,x_5=0.}
This describes for every $\theta$ an $\RP^4$, which coincides at $\theta=0$
or $\theta=\pi$ with the original $\RP^4$ defined in \ikx.  But starting
with a given twisted orientation at $\theta=0$, one comes back at $\theta=\pi$
with the opposite twisted orientation.  One way to show this is that,
as $\RP^5$ is orientable, to give a twisted orientation to $\RP^4$ is
the same as giving a twisted orientation of its normal bundle.  This
is concretely a one-form, nonzero in the direction normal to $\RP^4$, that
is odd under sign change of all $x_i$.  The one-form $\cos\theta\, dx_6
+\sin\theta\,dx_5$ does the job.  It obviously is continuous in $\theta$
and odd under sign reversal of the $x_i$, 
and has opposite sign at $\theta=\pi$ relative to $\theta=0$, confirming
that the twisted orientation is reversed in going around this path.

Now we want to count the elementary strings produced in the annihilation of
the two incoming fat strings.   The reason that such elementary strings
are created is somewhat similar to the reason that the wrapped fivebrane
studied in section  2 behaves as a baryon vertex.  The key ingredient in
section 2 was that fiveform flux integrated over the fivebrane contributes
to the charge that couples to the $U(1)$ field on the fivebrane.
In the present context, this means that the fiveform flux integrated over
$Z$ equals the total violation of chirality in the annihilation of the two
fat strings, and hence the number of elementary strings produced, modulo two.
We recall that from group theory, this number should be $N/2$, modulo two.

To compute the total charge violation, we note that the map from
$Z$ to $\RP^5$ is generically one-to-one.
(This is so because for a generic value of $(x_1,x_2,\dots,x_6)\in
\RP^5$, the equation $\cos\theta\,x_6+\sin\theta\,x_5=0$ is obeyed
for a unique value of $\theta$ with $0\leq\theta\leq \pi$.)  So the flux
integral over $Z$ equals that on $\RP^5$.  On $\RP^5$
there are $N/2$ units of five-form flux ($N$ on the covering space $\S^5$).
So chirality is violated by $N/2$ units, modulo two, in agreement with
expectations from gauge theory.\foot{
A more rigorous version of this discussion could be given using
mod two cohomology instead of differential forms and would count the
chirality violation modulo two.}

\bigskip\noindent{\it BPS Property}

As a prelude to discussing the BPS properties of elementary and fat
strings,  let us consider ten-dimensional supersymmetric Yang-Mills theory.
The supersymmetry transformation law for the gauge field $A$ is
\eqn\hxz{\delta A_i=\epsilon^\alpha\Gamma_{i\,\alpha\beta} \lambda^\beta,}
with $\lambda$ the gluino, $\epsilon$ an anticommuting parameter, $\Gamma$
a gamma matrix, and $\alpha,\beta$ spinor indices of $SO(1,9)$.
This transformation law shows that if $n$ is a null vector, then
$n\cdot A $ is invariant under eight supersymmetries, namely those
associated with parameters $\epsilon$ such that 
$\epsilon^\alpha n^i\Gamma_{i\,\alpha\beta}=0$.  It follows that
if $C$ is a lightlike straight line in $\R^{10}$, then 
\eqn\polj{\Tr_R P\exp\int_CA}
is invariant under eight    supersymmetries, for any representation $R$ of 
the gauge group.

Ten-dimensional super Yang-Mills can be dimensionally reduced
to four-dimensional super Yang-Mills with ${\cal N}=4$ supersymmetry.
Four components of the ten-dimensional gauge field reduce to the 
four-dimensional gauge field, which we will still call $A$, and the
other six components become scalars $\phi_i,\,i=1,\dots,6$ in the adjoint
representation.  Let now $D$ be a spacelike straight line in
$\R^4$, and let $\vec m$ be a unit six-vector, given in  components by
$m_i,\,i=1,\dots, 6$ with $\sum_im_i^2=1$.  A four-dimensional
analog of the statement that \polj\ is invariant under
eight supersymmetries is the statement that 
\eqn\golj{\Tr_R P\exp\int_D(A+i\vec m\cdot \vec \phi)}
is invariant under eight supersymmetries (plus, in fact, eight
superconformal symmetries).  In essence, \golj\ is the dimensional reduction
of \polj, using a complex null vector whose components are the unit
tangent vector to $D$ together with $i\vec m$.

To make contact with an $AdS$ description, it is convenient to work with
Euclidean signature and to add to
$\R^4$ a point at infinity, making $\S^4$ -- the boundary of $AdS$ space.
Including the point at infinity, $D$ becomes
a great circle on $\S^4$.  What BPS configuration in $AdS$ space
corresponds to a Wilson line that wraps around $D$?  $D$ is the boundary
of an $AdS_2$ subspace of $AdS_5$.\foot{
In $\R^5$ with coordinates $y_1,\dots,y_5$ and $|\vec y|=\sqrt{\sum_iy_i^2}$,
one can regard $\S^4$ as the
space $|\vec y|=1$, and $AdS_5$
as the space $|\vec y|<1$ with metric $4 d\vec y^2/(1-|\vec y|^2)^2$.
One can take for $D$ the great circle in $\S^4$
given by the equations $y_3=y_4=y_5=0$,
and for $AdS_2$ the subspace of $AdS_5$ defined by the same equations.}
To make a BPS state invariant under the symmetries that preserve the
$AdS_2$, we need a brane   on $AdS_5\times \S^5$ or $AdS_5\times \RP^5$
whose worldvolume will be this $AdS_2$ times a suitable
submanifold of $\S^5$
or $\RP^5$.

If $R$ is the fundamental representation of $SU(N)$, $SO(N)$, or $Sp(N)$,
then in the proposal of \refs{\malda,\reyyee}, one simply uses an elementary
Type IIB string, placed at the point $\vec m$ of $\S^5$ or $\RP^5$.
The worldsheet of the BPS configuration is $AdS_2\times \{\vec m\}$.
What do we do if $R$ is the spinor representation of $SO(N)$?   In
this case, we must consider a fat string, whose worldvolume will be
$AdS_2\times \RP^4$, with some $\RP^4$ subspace of $\RP^5$.  
To make it possible to interpret the BPS operator \golj\ via $AdS$
fat strings, the choice of
a unit vector $\vec m$ must determine a particular $\RP^4\subset\RP^5$.
Happily, it does: the $\RP^4$ in question is simply the subspace 
of $\RP^5$ that is ``orthogonal'' to $\vec m$.  (In other words,
this $\RP^4$ is parametrized by $x_i$, $i=1,\dots,6$, defined
up to overall sign, with $\sum_ix_i^2=1$ and $\sum_im_ix_i=0$.)
This relies on the fact that we built the fat string by wrapping
precisely on a codimension one subspace of $\RP^5$; the codimension
one property is a non-trivial check, since it
was determined on grounds (namely, the dimensions
of fivebranes and of $\RP^5$) that are seemingly unrelated to the BPS
properties of loop operators.

\subsec{Domain Walls}

In this subsection, we consider objects in $AdS_5\times \S^5$
and $AdS_5\times \RP^5$ that look like threebranes in the five noncompact
dimensions of $AdS_5$.  In $AdS_5\times \S^5$, the only such object
is the unwrapped Type IIB threebrane.  In $AdS_5\times \RP^5$, in
addition to the unwrapped Type IIB threebrane, we have threebranes
made by wrapping a Type IIB fivebrane on $\RP^2\subset \RP^5$.

$AdS_5$ has four spatial dimensions, so a threebrane has codimension
one and could potentially behave as a domain wall, with the string
theory vacuum ``jumping'' as one crosses the threebrane.  We will
see that all threebranes mentioned in the last paragraph are domain walls
in that sense.

\bigskip
\centerline{\vbox{\hsize=4in\tenpoint
\centerline{\psfig{figure=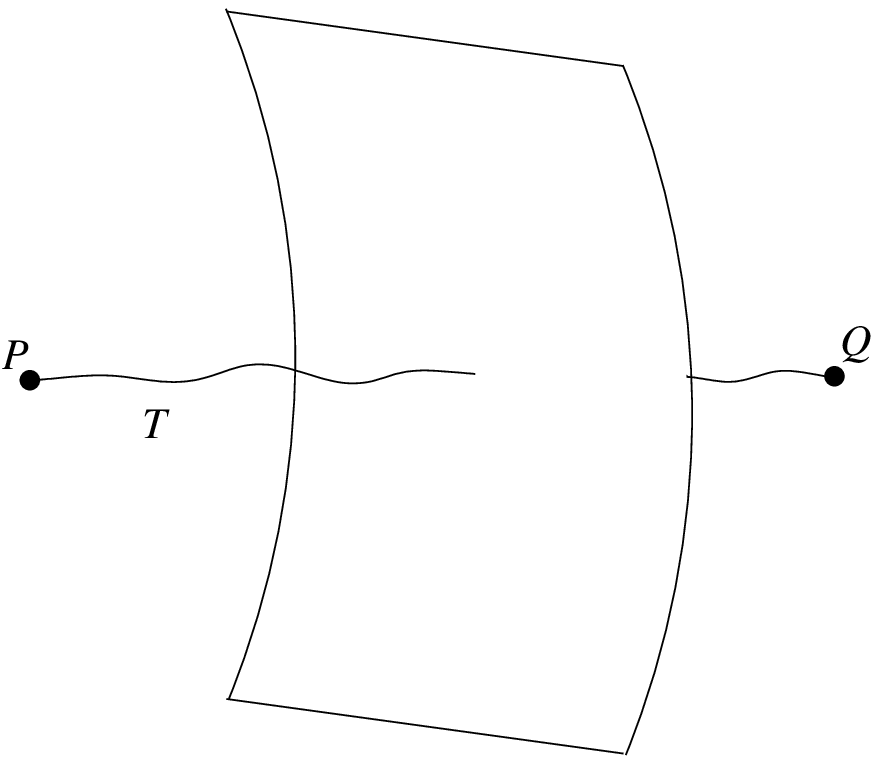}}
\vglue.4in
Fig. 7.  A path $T$ connecting two points $P$ and $Q$ that are on
opposite sides of a domain wall. }}

We begin by considering $AdS_5\times \S^5$.  The only threebrane
is the unwrapped Type IIB threebrane.  It is a source of the five-form
field $G_5$ of Type IIB string theory.  From this, it can be deduced
that in crossing the threebrane, the flux of $G_5$ over $\S^5$ jumps
by one unit.
The argument runs as follows.
Let $P$ and $Q$ be points on opposite sides of the threebrane, as in figure 7.
Let $T$ be a path from $P$ to $Q$, intersecting the threebrane once.
The six-manifold  $T\times \S^5$ intersects the threebrane at a single
point.
We consider the integral
\eqn\mumbo{\int_{T\times \S^5}{dG_5\over 2\pi}.}
This integral equals 1, since  $dG_5/2\pi$ is a delta function
supported on the threebrane, and as we have noted, the threebrane
intersects $T\times \S^5$ at one point.  On the other hand, we can evaluate
the integral using Stokes's theorem.  Since the boundary of $T\times \S^5$
is $P\times \S^5-Q\times\S^5$ (where the minus sign keeps track of the
relative orientation), we get
\eqn\klumbo{\int_{P\times \S^5}{G_5\over 2\pi}-
\int_{Q\times \S^5}{G_5\over 2\pi}  =1.}
This establishes the claim that the integrated five-form flux jumps
by one unit when one crosses the threebrane.

Thus the vacuum is different on the two sides.
The difference is easy to describe intuitively: the gauge group of the
boundary conformal field theory is determined by the five-form flux,
so it is $SU(N)$ on one side, and $SU(N\pm 1)$
on the other side.

\bigskip\noindent{\it Domain Walls In $AdS_5\times \RP^5$}

Now we consider the somewhat more subtle case that $\S^5$ is replaced
by $\RP^5$.  First we consider the unwrapped threebrane.  This can
be treated just as before.  The flux integral over $\RP^5$
\eqn\jccx{\int_{\RP^5}{G_5\over 2\pi}}
changes by 1 in crossing the threebrane.  So on the double cover
$\S^5$ of $\RP^5$, the flux integral
 changes by two.  Hence the gauge group jumps
in crossing the threebrane
from $SO(N)$ to $SO(N\pm 2)$, or from $Sp(k)$
(with $k=N/2$) to $Sp(k\pm 1)$.

Formally similar, but more subtle because torsion is involved, is
a Dirichlet or Neveu-Schwarz fivebrane wrapped on $\RP^2\subset \RP^5$
to make a threebrane.  
In this case, 
let $X$ be the four-manifold $X=T\times \RP^3$, with $T$ the same
curve as before, and $\RP^3$ a generic $\RP^3\subset \RP^5$.
Since a generic $\RP^2$
and $\RP^3$ in $\RP^5$ have one point of intersection, $X$ generically
intersects the fivebrane at one point.
 The boundary of $X$ is the union of the three-manifolds
$P\times \RP^3$ and $Q\times \RP^3$.  Because the fivebrane is a magnetic
source for the $B$-field and intersects $X$ in one point, the total
``magnetic charge'' of the $B$-field on the boundary of $X$ is non-zero.
If the $B$-field is trivial topologically on $P\times \RP^3$, it is nontrivial
on $Q\times \RP^3$, and vice-versa.  Thus, the discrete
``theta angle'' jumps in crossing the threebrane.

Which $\theta$ angle jumps depends on which fivebrane one considers.
By wrapping a Dirichlet fivebrane on $\RP^2$, we get a domain wall across
which $\theta_{RR}$ jumps; by wrapping an NS fivebrane on $\RP^2$,
we get a domain wall across which $\theta_{NS}$ jumps.  
Most surprising, from a conventional field theory point of view,
is the domain wall with a jump in $\theta_{NS}$; 
the boundary conformal field theory has orthogonal gauge group on
one side of this domain wall, and symplectic gauge group on the other.

An interesting property of domain walls made from fivebranes is that they
can carry threebrane charge.  We recall that on the fivebrane worldvolume $X$,
there is a $U(1)$ gauge field $a$; its field strength is a twisted two-form
$f$.  The topology of the $U(1)$ gauge field is determined by a characteristic
class $[f]\in H^2(X,\TZ)$.  As is usual in brane theory, the fivebrane
carries threebrane charge proportional to $[f]$.
Because $H^2(\RP^2,\TZ)=\Z$, a fivebrane wrapped on $\RP^2$ can carry
arbitrary threebrane charge; that is, it can absorb any number of
unwrapped threebranes.   This gives domain walls
across which the gauge group jumps from $Sp(k)$ to $Sp(k')$ with
arbitrary $k$ and $k'$, or similarly from $SO(2k)$ to $SO(2k'+1)$ or
$Sp(k')$.

\bigskip\noindent{\it Comparison To Flat Space Orientifold}

Like the fat strings of
section 4.2, the domain walls made by wrapping fivebranes on $\RP^2$
can be conveniently studied by going back to the flat-space orientifold
whose near-horizon geometry is $AdS_5\times \RP^5$.
\foot{As far as I know, the other main
examples in the present paper, which are the baryon vertex and the
Pfaffian particle, cannot be studied in a similar way.  
The stability of those objects
 depends on gravitational corrections to the flat-space orientifold.}
By doing so, we can help clarify an interesting phenomenon
found \ref\evans{N. Evans, C. V. Johnson, and A. D. Shapere,
``Orientifolds, Branes, And Duality Of 4D Gauge Theories,'' hep-th/9703210.}
in applications of orientifolds to gauge theory.
Hence in this discussion, we will consider orientifold $k$-planes,
not just three-planes.

Consider in $\R^{10}$, with coordinates $x^0,x^1,\dots,x^9$, a
$\Z_2$ transformation that leaves fixed $x^0,\dots,x^k$ and reverses
the sign of the others.  
The fixed point set is called an orientifold $k$-plane (in Type IIA or
Type IIB string theory for even or odd $k$).
The space normal to the $k$-plane looks like $\R^{9-k}/\Z_2$.  If we want
to analyze the behavior of string theory in this space using only
conventional geometry, we should keep away from the singularity at the origin.
The exterior to the singularity is contractible to $\RP^{8-k}$.
The exterior is in fact $\R^+\times \RP^{8-k}$, where $\R^+$ is parametrized
by the distance from the origin.

Since $H^3(\RP^{8-k},\TZ)=\Z_2$, there is a possibility of a discrete theta
angle in the exterior space.  As we know from our discussion in section
3, turning on $\theta_{NS}$ has the effect precisely of reversing the
sign of the elementary string amplitude for worldsheets of topology 
$\RP^2$ (and $\theta_{RR}$ is similarly related to $D$-strings).

In perturbative string theory, 
there are two types of orientifold $k$-planes.  They differ by the sign
of the orientifold projection for open strings -- the two choices lead
to orthogonal and symplectic gauge groups -- and by the sign of the
$\RP^2$ path integral.  Because of the last assertion (in relation to the
statement in the previous paragraph), the two types of $k$-plane differ
by the value of $\theta_{NS}$ in the smooth manifold exterior to the $k$-plane.
Via this interpretation, the two types of perturbative orientifold plane
can be distinguished just by measurements outside the plane -- though
to observe and distinguish the gauge group directly takes measurements
near the orientifold singularity.  (In the case $k=3$, if
one considers also $D$-strings,
one can make a further refinement and distinguish orientifold
planes by the value of $\theta_{RR}$ in the exterior region.)

Now consider an NS fivebrane whose world-volume is parametrized
by $x^0,\dots,x^{k-1}$ and by $x^{k+1},\dots,x^6$.  The intersection
of the fivebrane with the orientifold plane is of codimension one in that
plane and divides it, potentially, into domain walls.  It was in fact
shown in \evans\ that in crossing such a fivebrane, the ``type'' of the
orientifold plane is reversed.  In our present language, this means
that the value of $\theta_{NS}$ jumps in crossing the fivebrane. 
The explanation is just as above (indeed, our previous discussion corresponds
to the case $k=3$, while $k=4$ was considered in \evans).
In the directions normal to the orientifold plane, the fivebrane
is wrapped on $\R^+\times \RP^{5-k}$.  The Poincar\'e dual of 
$\RP^{5-k}$ in $\RP^{8-k}$  is the generator of $H^3(\RP^{8-k},\TZ)$.
The magnetic coupling of the $B$-field to the fivebrane means that in crossing
the fivebrane, the characteristic class of the 
$B$-field changes by this class, or in other words $\theta_{NS}$ jumps.

\subsec{The Baryon Vertex}

Here we will study the baryon vertex of orthogonal and symplectic
gauge theory -- adapting to $\RP^5$ the considerations of section 2
in the $\S^5$ case.

By analogy with section 2, one expects at first sight that the baryon
vertex will consist of a fivebrane wrapped once on $\RP^5$ -- say
a Dirichlet fivebrane if one wishes a baryon vertex connecting
external quarks of the electric gauge group.  Thinking along those
lines, one quickly comes to a paradox.  Suppose, for example, that
we are doing $SO(2k)$ gauge theory.  Then there are $k$ units of five-form
flux on $\RP^5$  ($2k$ units on the covering space $\S^5$).
Assuming that the fivebrane wraps once on $\RP^5$ and
following the reasoning of section 2, we then find that $k$ elementary
strings terminate on the fivebrane, and that those strings behave as fermions.
We thus seem to obtain a ``baryon vertex'' with an antisymmetric
coupling of  $k$ external
quarks.  But there is no gauge-invariant antisymmetric
combination, in $SO(2k)$ gauge theory, of $k$ external quarks.  The
baryon vertex of $SO(2k)$ gauge theory should couple $2k$ external quarks,
not $k$ of them.

What saves the day is that a state consisting of a fivebrane wrapped
once on $\RP^5$ does not exist.  Let $x$ be the generator of $H^1(\RP^5,\Z_2)$
and $X$ the fivebrane world-volume.  Let also $\Phi$ be the map of
$X$ to $AdS_5\times \RP^5$ given by the embedding\foot{Actually,
fivebranes of the topologies considered in the present section
cannot be embedded in $AdS_5\times \RP^5$.  But they can be mapped
to $AdS_5\times \RP^5$ via maps that are embeddings except in codimension
four.  (One does this by letting the brane ``wiggle'' generically
in $AdS_5$ while  wrapping on $\RP^5$.)     At points where the fivebrane
is not an embedded submanifold, there will be low energy modes that
cannot be seen using a long-wavelength fivebrane effective action; but
the codimension of the singularities is too high for such phenomena
to be relevant for us.}  of the fivebrane
in space-time.  As we discussed in section 3.3, we are limited to $X$
and $\Phi$ such that $\Phi^*(x)=w_1(X)$.

For most of the present section, we can ignore the
$AdS_5$ factor, because it is contractible, and consider $\Phi$ as
a map to $\RP^5$.  For a static fivebrane,  one has $X=Y\times\R$, 
with $\R$ the ``time'' direction and $Y$ a five-manifold.    This is 
contractible to $Y$, so topologically
we can think of $\Phi$ as a map of $Y$ to $\RP^5$.
For a static fivebrane wrapped once on $\RP^5$, we have $Y=\RP^5$
and $\Phi$ the identity map.  In this case, the condition $\Phi^*(x)=w_1(Y)$
is not obeyed, since, as $\RP^5$ is orientable, $w_1(\RP^5)=0$, while
for $\Phi$ the identity map, $\Phi^*(x)=x\not= 0$.  This at least
shows that we cannot get a contradiction by taking the fivebrane
worldvolume to be $\RP^5$.

To show  more generally that, regardless of the fivebrane topology, there is
no baryon vertex coupling $sk$ quarks for any odd integer $s$, we
want to show that if $Y$ is any closed five-manifold, and $\Phi:Y\to
\RP^5$ any map that obeys $w_1(Y)=\Phi^*(x)$, 
then $\Phi(Y)$ wraps an even number of times around
$\RP^5$.  This follows from the fact that for any closed five-manifold
$Y$, $w_1(Y)^5=0$.\foot{A proof using the Adem relations for Steenrod
squares was provided by D. Freed.  Using the fact that $Sq^k(w)=w^2$
for $w$ a $k$-dimensional class, and that $Sq^1$, as a map to the top
dimension, is the cup product with $w_1$, one has $w_1(Y)^5
=Sq^1w_1(Y)^4=Sq^1Sq^2w_1(Y)^2=Sq^1Sq^2Sq^1w_1(Y)=Sq^3Sq^1w_1(Y)$.
In the last step, one of the Adem relations was used.  But $Sq^3Sq^1w_1(Y)
=0$, since $Sq^r$ annihilates an $s$-dimensional class for $r>s$.}  
Hence if $w_1(Y)=\Phi^*(x)$,
one has $0=w_1(Y)^5=(\Phi^*(x))^5=\Phi^*(x^5)$.  But $x^5$ is the
mod two fundamental class of $\RP^5$, and maps $\Phi:Y\to \RP^5$ 
with $\Phi^*(x^5)=0$ are precisely those of even degree.

The basic non-trivial case is that $Y$ wraps twice around $\RP^5$.
As a simple example, we take $Y=\S^5$, with the natural two-to-one projection
to $\RP^5$.\foot{In other words, we regard $\S^5$ as the sphere
$\sum_{i=1}^6x_i^2=1$, and $\RP^5$ as the quotient of $\S^5$ by
$x_i\to -x_i$.  The ``identity'' map on the $x_i$ gives the degree
two map of $\S^5$ to $\RP^5$.}    
Since the degree of the map is two, the five-form field integrates
over $Y$ to twice the value on $\RP^5$.  Hence the problem discussed above
is avoided.   As desired, $Y$
gives an antisymmetric coupling of $N$  
external quarks, not $N/2$ of them, for $SO(N)$
or $Sp(N/2)$ gauge theory.

At this stage, however, we meet the following perplexing question.
$Y$ vanishes as an element of $H_5(\RP^5,\TZ)$, since that group is
in fact zero.  So why is the baryon vertex just found stable?
Before trying to discuss this question in the case of $AdS$ string theory, 
we will first review the situation in field theory.

\bigskip\noindent{\it Existence And Stability Of Baryon Vertex In Field Theory}

In $SO(N)$ or $Sp(N/2)$ gauge theory,  there is a fundamental representation
of dimension $N$.  The $N$-fold completely antisymmetric tensor
product of this representation is gauge-invariant.
If $\psi$ is a fermion valued in the fundamental representation,
this antisymmetric invariant is
\eqn\micci{B={1\over N!}
\epsilon_{i_1i_2\dots i_N}\psi^{i_1}\psi^{i_2}\dots \psi^{i_N}.}

This is the ``baryonic'' combination of $N$ external quarks, which
we have aimed to reproduce in $AdS$ space via the ``baryon vertex,''
at which $N$ elementary strings can join.  
Note that the invariant $B$ can be defined whether the gauge group is
$SO(2k)$, $SO(2k+1)$, or $Sp(k)$.  So a baryon vertex should exist
for each of the possible groups.

This is in agreement with what we have found above.  The manifold
$Y=\S^5$ has $H^3(Y,\TZ)=0$, so the requirement that the $B$-fields should
be topologically trivial when pulled back to $Y$ is automatically obeyed.
The use of $Y$ for a baryon vertex is equally valid for any value of 
$\theta_{NS}$ or $\theta_{RR}$.  The $AdS$
baryon vertex thus exists regardless of the gauge group of the boundary
theory.

\bigskip
\centerline{\vbox{\hsize=4in\tenpoint
\centerline{\psfig{figure=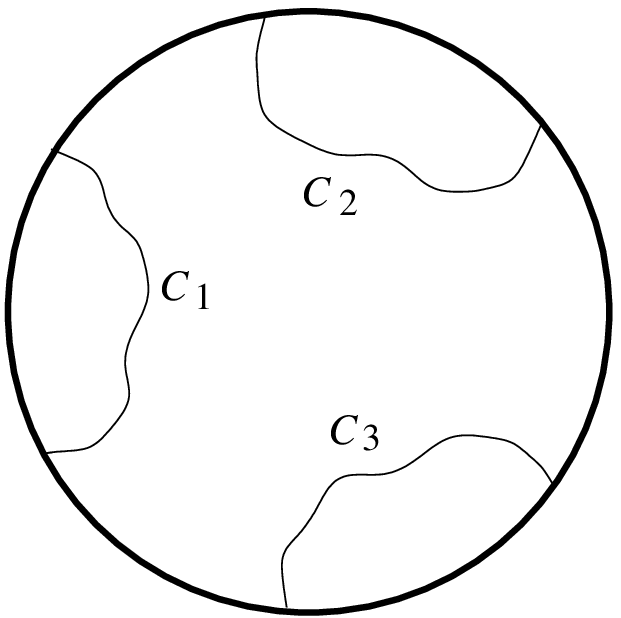}}
\vglue.4in
Fig. 8.  In $Sp(k)$ gauge theory -- as sketched here for $k=3$ -- a baryon
vertex can decay to a configuration in which $2k$ charges on the boundary
are connected pairwise by elementary strings $C_1,\dots,C_k$.}}

What about stability of the baryon in field theory? This is more
complicated. In $Sp(N/2)$
gauge theory, there is an invariant second rank antisymmetric tensor
$\gamma_{ij}$, via which one can form the ``meson'' $M=\half\gamma_{ij}
\psi^i\psi^j$.  A ``baryon'' of $Sp(N/2)$ can decay to mesons
since in fact
\eqn\hicci{B={1\over (N/2)!}M^{N/2}.}
Thus, we should expect no topological stability for the $AdS$ baryon
vertex when $\theta_{NS}\not= 0$.   For $\theta_{NS}\not= 0$,  an initial
state with a fivebrane wrapped twice on $\RP^5$ and connected by $N$
elementary strings to charges on the boundary should, topologically, be able
to decay to a state (indicated in figure 8)
with no fivebrane and with $N/2$ strings that
join the external quarks pairwise.  (This decay is not necessarily
favored energetically.)  

The case of $SO(N)$ is more subtle.  The analog of 
$\gamma_{ij}$ is the ``metric,'' the symmetric tensor $\delta_{ij}$.
We must take account of the fact -- already used in our discussion of
Pfaffians in section 4.1 -- that $\N=4$ super Yang-Mills theory
with gauge group $SO(N)$ actually has $O(N)$  symmetry, not just
$SO(N)$.  The generator $\tau$ of the quotient $O(N)/SO(N)$ behaves
as a global symmetry.  $\delta_{ij}$ is invariant under 
$O(N)$, while $\epsilon_{i_1i_2\dots i_N}$ is odd under $\tau$.
So the baryon --  which is odd under $\tau$ -- cannot decay to mesons -- 
which are even
under $\tau$.  Hence, the transition sketched in figure 8 should
be impossible for $SO(N)$, that is, for $\theta_{NS}=0$.

On the other hand, in $SO(N)$ gauge theory there are states odd
under $\tau$ other than the baryon.  For $N$  even, one has the ``Pfaffian''
combination of $N/2$ gauge bosons, which we interpreted in section 4.1
in terms of a wrapped three-brane.  For $N$ odd, the wrapped three-brane
joined by a string to an external charge similarly represents
a $\tau$-odd state.

\bigskip
\centerline{\vbox{\hsize=4in\tenpoint
\centerline{\psfig{figure=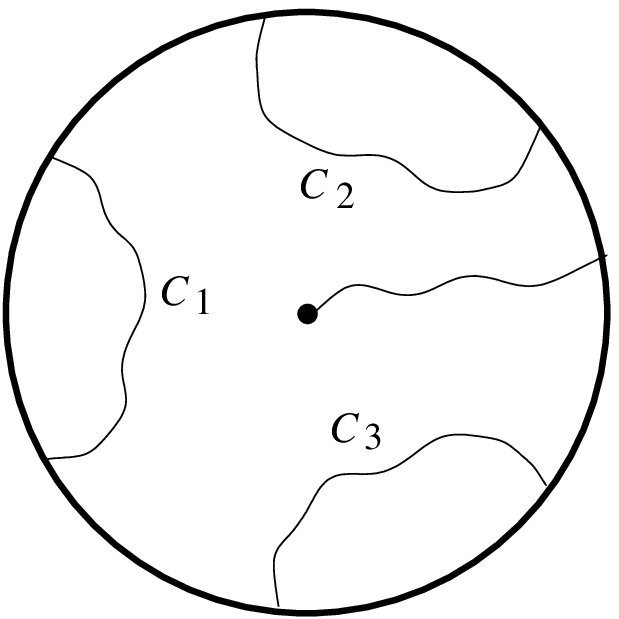}}
\vglue.8in
Fig. 9.  In $SO(2k)$ gauge theory, the baryon vertex can decay to
a state containing a wrapped threebrane plus strings making pairwise
connections between external charges.
In $SO(2k+1)$, which is the case sketched here for $k=3$, 
the final state contains
in addition an odd number of strings connecting the wrapped threebrane
-- indicated as a heavy dot in the interior -- to the boundary. In the
figure, there is one such string.}}

For $SO(N)$, decay to elementary strings only should be forbidden, but
there should be processes in which 
a state containing 
a baryon vertex constructed from a wrapped fivebrane is transformed
to a state containing a wrapped threebrane together
with elementary strings, as
sketched in figure 9.  Both are odd under $\tau$, so  transitions between them
should be possible.
After analyzing, albeit in an incomplete fashion, the decay
 of the baryon vertex to elementary strings, we will briefly discuss
 the decay to a state containing a wrapped threebrane.

\bigskip\noindent{\it Decay Of Baryon Vertex In $AdS$}

To interpret the decay of a baryon to mesons
 in $AdS$ space, we proceed as follows.
We consider a state that up to time $t_0$ has a wrapped fivebrane with
topology $\S^5$.  Thus, this part of the fivebrane world-volume 
has topology $\R_-\times \S^5$, where $\R_-$ is the part of the ``time''
line with $t\leq t_0$.  To describe the decay of the fivebrane, we find
a compact six-manifold $X$, with boundary $\S^5$,  and 
we take the total fivebrane worldvolume $Y$
to be the manifold obtained by gluing $\R_-\times \S^5$ onto $X$ along their
common boundary.
The map from $Y$ to spacetime is specified by picking a map
$\Phi:X\to AdS_5\times \RP^5$ that maps the boundary to $\{t_0\}\times \S^5$.
  Thus, the fivebrane worldvolume comes
in from the far past, and ``ends'' within a finite time of $t_0$.
(What the attached strings are doing meanwhile will be discussed later.)

The map $\Phi$, which topologically can be viewed as a map of $X$ to $\RP^5$,
must obey the usual condition $\Phi^*(x)=w_1(X)$, and must agree
on the boundary of $X$ with the usual two-to-one projection from $\S^5$ to
$\RP^5$.  (The last condition ensures that the maps from $\R_-\times \S^5$ and
from $X$ to spacetime
glue correctly to a map from $Y$.)     $\S^5$ with the two-to-one map to $\RP^5$
 vanishes
as an element of $H_5(\RP^5,\TZ)$ (as that group is actually zero),
so there exists a manifold $X$ and map $\Phi$ obeying 
the given  conditions.  An explicit example is as follows.
In $\R^7$, with coordinates $x_1,\dots,x_7$, take the subspace
$\S^5\times I$ ($I$ is the unit interval) defined by $\sum_{i=1}^6x_i^2=1$,
$|x_7|\leq 1$.  Divide $\S^5\times I$ by the $\Z_2$ transformation
$x_i\to -x_i,\,i=1,\dots,7$;
let $X$ be the quotient.  By forgetting  $x_7$, $X$ maps to $\RP^5$;
this is the desired map $\Phi$.  The boundary of $X$ is the double cover
of $\RP^5$ given by taking $x_7=\pm 1$.  This double cover is $\S^5$
(since one can divide $X=\S^5\times \{x_7=\pm 1\}$ by $\Z_2$ by restricting
$x_7$ to be $+1$), and $\Phi$ induces on $\S^5$ the usual two-to-one
projection to $\RP^5$.  
This shows that $X$ and $\Phi$ have the desired properties.

So we have found a mechanism for the decay of the fivebrane.
However, we must impose the usual condition on topological triviality
of the NS $B$-field when pulled back to a Dirichlet fivebrane worldvolume.
For the baryon vertex itself, this was no problem, as $H^3(\S^5,\TZ)=0$.
However, $H^3(X,\TZ)$ is non-trivial, so there is a potential obstruction
to decay of the baryon vertex via the manifold $X$.  In fact,
$X$ is contractible to $\RP^5$ (by squeezing the $x_7$ axis down to zero),
so $H^3(X,\TZ)$ is naturally isomorphic to $H^3(\RP^5,\TZ)=\Z_2$.  The map
$\Phi:X\to \RP^5$ is actually a homotopy equivalence, and $\Phi^*$ is
therefore an isomorphism.  

If $[H]$ is the characteristic class of the NS $B$-field, then the
condition stated in section 3.3, namely $i^*([H])=0$, implies that
$[H]=0$.  This means that the decay of the baryon vertex by this mechanism
is possible if and only $\theta_{NS}=0$.  The decay would, in other
words, occur for orthogonal and not for symplectic gauge groups, which
is precisely the wrong answer!  This suggests that we should look for
an ``overall minus sign'' that will exchange the two cases.
It was, in fact, suggested at the end
of section 3.3 that the general condition is really $i^*([H])=W$,
with $W$ a certain natural element of $H^3(X,\TZ)$.  It can be shown\foot{The
normal bundle to $\RP^5\subset X$ is the unorientable real line bundle
over $\RP^5$; also, $X$ is homotopic to $\RP^5$, so one can evaluate
$W$ by restricting to $X$.  The total Stieffel-Whitney class of $X$, taking
account the normal bundle to $\RP^5$, is $(1+x)^7$, where $x$ is
the generator of $H^2(\RP^5,\Z)=\Z_2$.  In particular, $w_2(X)=x$ and is
non-zero.
Now consider the long exact cohomology sequence derived from \iloo.
$w_2(X)$ cannot be lifted to a class in $H^2(\RP^5,\TZ)$ (as that group
vanishes), so $W\not= 0$ and hence generates $H^3(X,\TZ)=\Z_2$.  For
all other cases in the present paper, $W=0$ since either $w_2(X)=0$ 
(branes wrapped on $\RP^3$ or $\RP^4$) or
$H^3(X,\Z)=0$ (branes wrapped on $\RP^2$).}
that for the manifold $X$, $W$ is the generator of $H^3(X,\TZ)$, while
$W=0$ for all other brane worldvolumes considered in the present paper.
Thus, if the proper condition is $i^*([H])=W$, then  brane decay
by the mechanism discussed here is actually possible if and only if
$\theta_{NS}\not=0$, that is, if and only if the gauge group is symplectic.
This is a strong hint that the proper condition involves the $W$ term,
a matter that is under investigation \frw.

It remains to discuss what happens to the strings while the fivebrane
is being capped off by the manifold $X$.  The baryon vertex on $\R_-\times
\S^5$ is connected to the boundary of $AdS_5$ by $N$ strings whose
worldsheets end in curves $C_i$ on $\R_-\times \S^5$.  One can take
these curves to be of the form $\R_-\times P_i$, with $P_i$ some
points in $\S^5$.  To complete the description of the decay of the baryon
vertex, the union of the $C_i$ must be extended to a collection of
curves without boundary on $Y$.  This must
be done by taking the $N$ boundary points of the $C_i$ on $\S^5$
(these are the points $t_0\times P_i$), and connecting them pairwise
via strings in $X$.  If $N$ is odd, this is impossible, as an odd number
of points cannot be joined pairwise.  This corresponds to the statement
that in $SO(N)$ gauge theory with odd $N$, a ``baryon'' cannot decay to mesons
simply because it contains an odd number of quarks; any decay of the baryon
vertex will, as we have seen, involve a wrapped threebrane in the final state.
For even $N$, however,
there is no obstruction to joining the ends of the $C_i$ pairwise
(because $X$ is unorientable, this can be done in a way that is compatible
with the twisted orientations of the $C_i$; one merely loops around an
orientation-reversing loop in $X$ whenever needed).  
Once the $C_i$ have been extended
over $Y$, the resulting curves can be
connected to the boundary of $AdS$ space via string worldsheets,
completing the description of the decay of the baryon vertex.

\bigskip\noindent{\it Decay To State Containing Wrapped Threebrane}

Now we will briefly analyze the decay of the baryon vertex to
a state containing a threebrane wrapped on $\RP^3$.  We recall from
our group theory discussion that such a decay should be possible
when the gauge group is $SO(N)$ -- which is the case that the baryon
vertex cannot decay to a state containing mesons only.  

The basic reason that this is possible is that threebranes can end on
fivebranes.  Moreover, the end of the threebrane on a fivebrane worldvolume
is a magnetic source for the $U(1)$ gauge field $a$ that propagates on the
fivebrane.  Let $X$ be the fivebrane worldvolume, $E$ the worldvolume
of a threebrane whose boundary is on $X$, and $D$ the boundary of $E$.
Then $D$ is an orientable three-manifold; 
it is orientable because the threebrane
worldvolume is always orientable, and the boundary of an orientable manifold
is orientable.  Consequently, the Poincar\'e dual of $D$ is a class
$[D]\in H^3(X,\TZ)$.  Because the equation\foot{In this discussion
we will have to assume again that the $W$ term is really present!}
 $i^*([H])=W$ is really
the Bianchi identity for $a$, and $D$ acts as a magnetic source for $a$,
the equation becomes in the presence of a threebrane
\eqn\ucci{i^*([H])=W+[D].}
The $[D]$ term here is just analogous to the $[C]$ term in \recallit,
which governs strings ending on threebranes.
To be more precise, \ucci\ holds for Dirichlet fivebranes with $[H]=
[H_{NS}]$, which is the case we will actually consider, or for
NS fivebranes with $[H]=[H_{RR}]$.

Now recall that the mechanism that prevents decay of the $SO(N)$
baryon vertex to strings only
is that when the gauge group is $SO(N)$, one has $i^*([H])=0$; but $W$
is not zero for the fivebrane worldvolume that describes decay of the baryon
vertex.  What happens if threebranes are included?
We see from \ucci\ that decay of the baryon vertex
to a state containing threebranes
as well as strings is possible if $W+[D]=0$ or equivalently
(since $W$ is a two-torsion class) $[D]=W$.

This condition can be obeyed and
corresponds, as expected, to having in the final state a single
threebrane (or an odd number of them) wrapped on $\RP^3\subset \RP^5$.
Indeed, as the fivebrane worldvolume $X$ is contractible to $\RP^5$, 
we can take it to contain a copy of $\RP^5$, say at some time $t_1$
and at some point $P$ in the spatial part of $AdS_5$.  Let $D$ be
any $\RP^3$ subspace of this $\RP^5$.  Then $[D]=W$ (each of them
being the nonzero element of $H^3(\RP^5,\TZ)$).   To give
a threebrane worldvolume $E$ with boundary $D$, we simply let $\R_+$ be
the product of the set in $AdS_5$ with $t\geq t_1$ and position $P$ in space,
and we take $E=\R_+\times \RP^3$.  Clearly, with this choice of $E$,
we have decay of the baryon vertex to a state containing a single wrapped
threebrane.  In the final state, the $N$ external charges are connected
to each other or to the threebrane by elementary strings.

\subsec{Instantons In $AdS$ Space}

We will here conclude with a brief observations about the one Type IIB
brane that we have so far overlooked -- the $-1$-brane.
It has already been noted \banks\ that $-1$-branes
should be identified with instantons of the boundary conformal field theory.
Here we will note an interesting fact relevant to this identification.

Consider the moduli space of $SU(2)$ instantons on $\S^4$ of instanton
number one.  Any such instanton is invariant under an $SO(5)$ subgroup of
the conformal group $SO(5,1)$ of $\S^4$ \ref\jackiw{R. Jackiw and C. 
Rebbi, ``Conformal Properties Of Pseudoparticle Configurations,''
Phys. Rev. {\bf D15} (1977) 1642.}.
Any two such instantons are related by an $SO(5,1)$ transformation.
The moduli space of such instantons is hence a copy of $SO(5,1)/SO(5)$,
that is a copy of $AdS_5$.  By contrast, the moduli space for a single
$-1$-brane on $AdS_5\times \S^5$ is, of course, just a copy of
$AdS_5\times \S^5$.  Clearly, a similar statement holds if $\S^5$
is replaced by $\RP^5/\Z_2$.   
It is tempting to identity the $AdS_5$ moduli space
of the instanton with the first factor in the moduli space of the
$-1$-brane; the relation between them hopefully comes by somehow averaging
over the $-1$-brane position on $\S^5$ or $\RP^5$.  

To consider the $k$-instanton moduli space, one should begin with
gauge group $SU(N)$ for some large $N$.  One component of the $k$-instanton
moduli space is described by placing the $k$ instantons in $k$ commuting
factors of $SU(2)$.  This component is a symmetric product of $k$ copies of
$AdS_5$.  The moduli space of the same number of
 $-1$-branes is meanwhile a symmetric
product of $k$ copies of $AdS_5\times \S^5$ or $AdS_5\times \RP^5$, 
obviously a closely related
answer.  The $k$-instanton moduli space also has other components, for instance
with all $k$ instantons in a common $SU(2)$.  Perhaps these components
make nonleading contributions for large $N$, in which case they might
be difficult to see in the $AdS$ description.

I should note in conclusion that a brane wrapping mode that has {\it not}
been interpreted in the present paper is the twobrane made by wrapping
a threebrane on $\RP^1\subset \RP^5$.  It would be interesting to know
its interpretation in the boundary conformal field theory.
\bigskip
This work was supported in part by NSF Grant PHY-9513835.  I would
like to thank J. Blum, D. Freed, and N. Seiberg for discussions and comments.
\listrefs
\end